\documentclass[final,5p,times,twocolumn]{elsarticle}
\usepackage{graphicx}
\usepackage{subcaption}
\usepackage{amsmath,amsthm,amsfonts,amssymb,amscd,epsfig, enumerate,yfonts}
\usepackage{tikz}
\usepackage{ragged2e}
\usepackage{booktabs}
\usepackage[numbers,sort&compress]{natbib}
\usepackage{float}
\usepackage{lineno}
\usepackage{color}
\usepackage{mathtools}
\usepackage{showkeys}
\usepackage{siunitx}
\usepackage{cuted}
\usepackage{multirow}

\usepackage[colorlinks=true,linkcolor=red,citecolor=cyan]{hyperref}
\modulolinenumbers[200]
\biboptions{sort&compress}

\journal{Journal Name}

\begin{document}

\begin{frontmatter}

\title{Gravitational wave radiation from periodic orbits and quasi-periodic oscillations in Einstein non-linear Maxwell-Yukawa black hole}

\author[address1]
{Tehreem Zahra}
\ead{tehreemzahra971@gmail.com}
\author[address1]
{Oreeda Shabbir}
\ead{oreedashabbir7@gmail.com}
\author[address3]
{Bushra Majeed}
\ead{bushra.majeed@ceme.nust.edu.pk}
\author[address1,address9]{Mubasher Jamil}
\ead{mjamil@sns.nust.edu.pk}
\author[address5,address8]
{Javlon Rayimbaev}
\ead{javlon@astrin.uz}
\author[addrAbu1,address6,addrAbu2]{Abubakir Shermatov}
\ead{shermatov.abubakir98@gmail.com}

\address[address1]{School of Natural Sciences, National University of Sciences and Technology (NUST), Islamabad, 44000, Pakistan}
\address[address3]{College of Electrical and Mechanical Engineering (CEME), National University of Sciences and Technology, Islamabad, 44000, Pakistan.}
\address[address9]{Research Center of Astrophysics and Cosmology, Khazar University, Baku, AZ 1096, 41 Mehseti Street, Azerbaijan}
\address[address5]{National University of Uzbekistan, Tashkent 100174, Uzbekistan}
\address[address8]{Urgench State University, Kh. Alimdjan Str. 14, Urgench 220100, Uzbekistan} 
\address[addrAbu1]{Institute of Fundamental and Applied Research, National Research University TIIAME, Kori Niyoziy 39, Tashkent 100000, Uzbekistan}
\address[address6]{University of Tashkent for Applied Sciences, Str. Gavhar 1, Tashkent 100149, Uzbekistan}
\address[addrAbu2]{Tashkent State Technical University, Tashkent 100095, Uzbekistan
}
\date{Received: date / Accepted: date}

\begin{abstract}
In this article, we investigate the orbital dynamics and quasi-periodic oscillations (QPOs) surrounding a static, spherically symmetric geometry of an Einstein–nonlinear Maxwell–Yukawa (ENMY) black hole (BH). Using the Hamiltonian formalism, we derive equations of motion and analyze the effective potential. We determine the innermost stable circular orbits (ISCO) and innermost bound circular orbits (IBCO) radii for different values of the Yukawa parameters $\lambda$ and $\delta$, and classify periodic orbits via rational frequency analysis, highlighting deviations from Schwarzschild geometry. We also study gravitational wave (GW) emission from periodic orbits and show how Yukawa terms affect GW signals. Fundamental frequencies are computed, and QPOs are analyzed using relativistic precession, warped disk, and tidal disruption models. By increasing $\lambda$, the ENLMY spacetime effectively mimics the behavior of a Schwarzschild spacetime. Constraints on the BH mass and Yukawa parameters are derived using QPO data from stellar-mass (XTE J1550-564, GRO J1655-40, GRS 1915+105), intermediate-mass (M82 X-1), and supermassive (SgrA*) BHs within the relativistic precession model by employing a Markov Chain Monte Carlo analysis.
\end{abstract}

\begin{keyword}
Einstein–Nonlinear Maxwell–Yukawa black hole, modified gravity, $f(R)$ gravity, effective potential, circular orbits, innermost stable circular orbits, innermost bound circular orbits, periodic orbits, gravitational waves, test particle dynamics, Yukawa correction, Hamiltonian formalism, geodesic motion, quasi-periodic oscillations
\end{keyword}

\end{frontmatter}

\linenumbers
\section{Introduction}
General Relativity (GR) has been remarkably successful in describing gravitational phenomena, from solar system tests to gravitational wave (GW) detections \cite{LightDeflection,Everitt:2011hp,GWs,EHT,shapiro1964fourth,pulsars}. However, it lacks a consistent quantum description \cite{QF} and fails at singularities. Additionally, GR faces challenges in describing the accelerated expansion of the universe without invoking unknown components, such as dark matter (DM) and dark energy (DE). There is no direct evidence for the existence of DM, and a theoretical framework to describe its behavior is still lacking \cite{DM1,DM2}. These limitations have motivated the development of modified theories of gravity, which can be done by either introducing new matter fields or by altering the geometric structure of spacetime \cite{modified1,modified2,modified3}. Among the latter is $f(R)$ gravity, which modifies the Einstein-Hilbert action by promoting the Ricci scalar $R$ to a more general function $f(R)$. Such modifications naturally introduce additional degrees of freedom and can provide alternative explanations for cosmological and galactic dynamics without the need for unknown matter components. In the weak-field limit, $f(R)$-gravity yields Yukawa-like corrections to the gravitational potential. Such a correction carries a characteristic length scale that can effectively screen gravitational modifications at solar system scales while revealing deviations in the strong-field regime \cite{yukawa3}. Hence, these extended theories must be thoroughly tested in astrophysical environments to determine their validity. In this context, black holes (BHs), as the most compact and extreme gravitational objects in the universe, provide a natural setting for testing gravitational theories in the strong-field regime \cite{bambi2017black}.
\par
The properties of geodesic orbits are deeply connected to the underlying BH geometry. Among these, periodic orbits play a crucial role in understanding orbital dynamics \cite{periodic}, both in astrophysical contexts, such as planetary motion in the Solar System, and more extreme environments, like galactic centers. In particular, periodic orbits are essential for analyzing the long-term stability of bound systems and offer valuable insights into the evolution of extreme mass-ratio inspirals (EMRIs) \cite{periodic1}. An EMRI system is formed by a stellar-mass object orbiting around a central supermassive BH \cite{emris}. During the adiabatic inspiral phase, periodic orbits serve as transitional paths that significantly influence the GWs emitted by such systems \cite{Babak:2006uv}. These waveforms are highly sensitive to the orbital configuration, making periodic geodesics vital for GW detection and signal interpretation. These orbits often exhibit zoom-whirl behavior, with the test particle returning to its initial location after a finite time \cite{shabbir2025periodic}. In \cite{PhysRevD.79.043016,PhysRevD.79.043017,babar2017periodic}, periodic orbits in Schwarzschild, Kerr, and naked singularity spacetimes are investigated extensively. In this work, we investigate timelike periodic geodesic orbits in the spacetime of a Yukawa-like BH. In addition to identifying such orbits, we examine the parameter space of angular momentum and energy that permits bound motion in this modified gravitational background, thereby understanding how deviations from GR affect orbital structure.
\par
Studying the spectroscopic properties of accretion disks around compact astrophysical objects (neutron stars, BHs, and hypothetical wormholes) through Fourier analysis of their X-ray emissions provides a powerful method for probing spacetime structure and revealing characteristic accretion processes. In X-ray binary systems (XRBs) with BHs or neutron stars and companion stars, the central object's gravity affects radiation processes in the accretion disk \cite{Stella1999ApJ}. Such analyses have identified quasi-periodic oscillations (QPOs), which are peaks observed in the power density spectrum of X-ray emissions \cite{vanderKlis:milliarsec}. These are categorized as high-frequency (HF, 50--450~Hz
) or low-frequency (LF, 0.01--30~Hz) QPOs \cite{Rizwan_2019}. Precise QPO measurements in microquasars and quasars offer insights into spacetime properties near BHs, enabling tests of gravity theories, constraints on BH parameters, and determination of the innermost stable circular orbits (ISCO) radius. 
\par
Previous studies \cite{JRqpo2021,Rayimbaev2021QPO,Rayimbaev2022CQGra,Rayimbaev2022EPJCEMSQPO,Rayimbaev2022IJMPDQPOcharged,2022PDU....3500930R,JR.galax.qpo,JR.qpoAP.23,Rayimbaev2023EPJC...83..572R} suggest that QPO orbits may help estimate the ISCO radius within observational uncertainties. Additionally, the capture of massless and massive particles by parameterized BHs has been examined \cite{2021Galax...9...65T,2021Univ....7..307A,2024PDU....4401483R}, while orbital and epicyclic frequencies in axially symmetric spacetimes are explored in \cite{2022Univ....8..507T,2021IJMPD..3050037T,2023Galax..11...70T}. Recent numerical studies have investigated QPO generation mechanisms by solving general relativistic hydrodynamic equations \cite{DONMEZ2006256} in Kerr and hairy metrics. These simulations reveal spiral shock waves formed by plasma perturbations in strong gravitational fields, linking them to QPO formation \cite{2024RAA....24h5001D,2024EPJC...84..524D,2024MPLA...3950076D}. Similarly, Bondi-Hoyle-Lyttleton accretion models describe shock cone dynamics and their role in producing QPO frequencies \cite{Koyuncu:2014nga,2022PhLB..82736997D,2024Univ...10..152D}. Such shocks may explain LF/HF QPOs in sources like GRS 1915+105 \cite{2024arXiv240810102D} and predict QPOs near the M87 BH \cite{2024arXiv240701478D}. However, the physical origin of QPOs is still unknown. Following the initial detection of QPOs, various theoretical models have been proposed to interpret their origin, including disko-seismic oscillations, relativistic precession (RP) model, tidal disruption (TD) model, the warped disk (WD) model, and resonance models \cite{2023EPJC...83..323K}. Using the RP model, constraints on the parameters of XRBs have also been found using statistical techniques in literature \cite{2014MNRAS.437.2554M,Bambi2013arXiv1312.2228B,liu2023constraints,Motta:XTE226,Motta:XTE564}. In the present work, we also calculate oscillation frequencies and apply them to QPOs to obtain constraints on the black hole and ENLMY gravity parameters using Markov Chain Monte Carlo (MCMC) analyses. 
\par
This paper explores the dynamics of test particles around an Einstein-nonlinear Maxwell-Yukawa (ENLMY) BH, and the analysis will be carried out by studying the periodic orbits and QPOs. We emphasize that the Yukawa corrections are not just mathematical artifacts but physical mechanisms of screening that allow the ENLMY geometry to interpolate between GR and modified gravity regimes. This feature makes ENLMY black holes particularly suitable for testing intermediate-scale deviations from GR in astrophysical environments. The outline of this paper is as follows: In Section 2, we introduce the ENLMY metric. In Section 3, the dynamics of test particles around an ENLMY BH is studied, including the effective potential and energy efficiency. Section 4 focuses on the effective potential and the conditions for circular orbits, including a detailed analysis of the ISCO and the innermost bound circular orbits (IBCO). Section 5 investigates the periodic orbits, analyzing their structure and behavior within the framework of the underlying spacetime geometry. In section 6, we examine the GW radiation emitted by periodic orbits in the ENLMY BH spacetime and analyze how Yukawa corrections influence the resulting waveforms. In Section 7, we find the fundamental frequencies of the test particle and carry out the analysis of QPOs phenomena using various precession models. Section 8 presents the constraints on BH mass and Yukawa-type parameters within the framework of the RP model. Finally, we will provide a brief conclusion to our results. Throughout the paper, we use the metric signature $(-,+,+,+)$ and units in which $G = c = 1$, unless stated. 

\section{The Einstein-Nonlinear Maxwell-Yukawa Metric}
\begin{figure}[ht]
    \centering
    \includegraphics[width=0.45\textwidth]{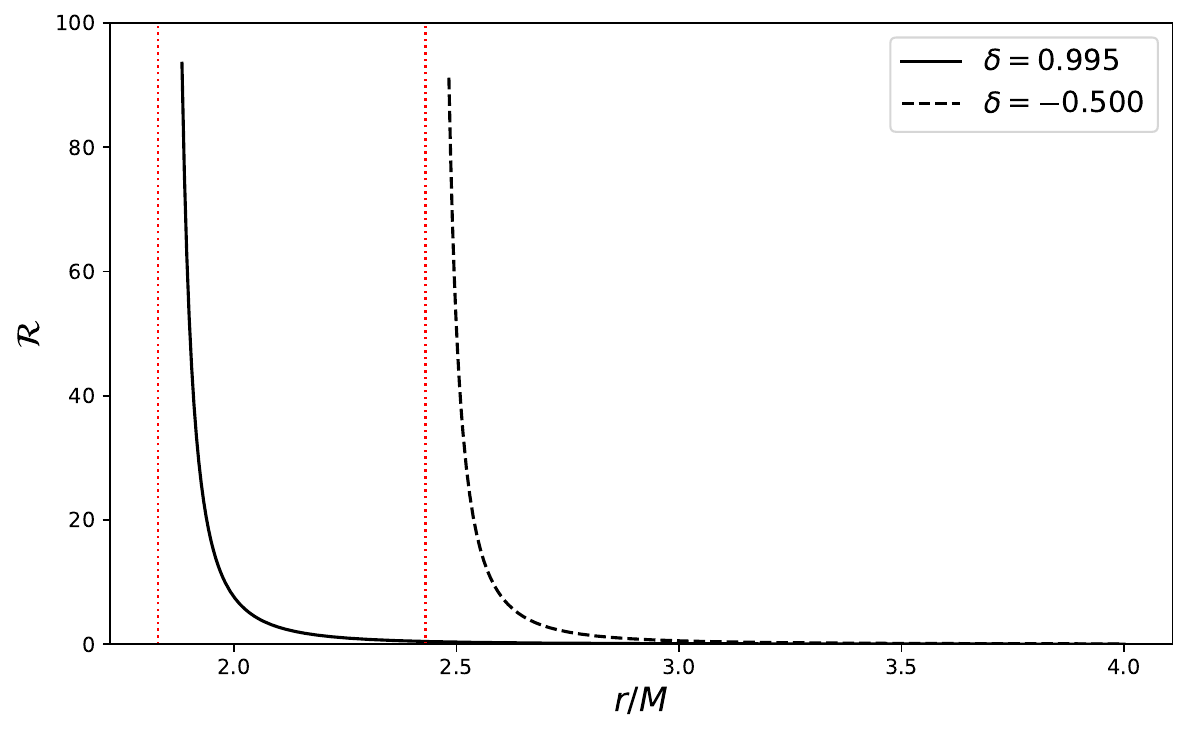}
    \caption{Ricci scalar as a function of the radial coordinate $r$, plotted for two values of $\delta$, with fixed $M = 1$ and $\lambda = 10$. Vertical red dotted lines mark the corresponding event horizons.}
    \label{fig:ricci}
\end{figure}

The ENLMY BH metric emerges from a spherically symmetric, static solution in $f(R)$ gravity, where the gravitational action is generalized by replacing the Ricci scalar $R$ with an analytic function $f(R)$
\begin{equation}
\mathcal{A}= \frac{c^4}{16 \pi G}\int d^4x \sqrt{-g}   f(R) + \mathcal{L}_m.
\end{equation}
Choosing the special $f(R)$, which can be expressed in the form of a Taylor series, expanded around a fixed point $R_0$.  
In the weak-field limit, a Taylor expansion of $f(R)$ and truncation at leading order yields modified field equations as \cite{yukawa4}
\begin{equation}
f(R)= {\Sigma}_n \frac{f^n (R_0)}{n!}(R-R_0)^n \approx f_0 +f'_0  R +\frac{f''_0}{2} R^2+..., 
\end{equation}
 where $f_0$ represents the cosmological constant, with $f'_0=1$. 
 This extended theory of gravity yields a solution characterized by a Yukawa-like potential, which represents a post-Newtonian correction to the Schwarzschild geometry \cite{yukawa3}.
This model is described by
the metric given by \cite{yukawaMetric,yukawa2}
\begin{eqnarray}\label{1}
ds^{2}&=&-[1+\Phi (r)] dt^2+ [1-\Phi (r)] dr^2 \nonumber \\
& & +r^2(d\theta^2 +\sin^2\theta
d\phi^2),
\end{eqnarray}
where
\begin{equation}
\Phi=-\frac{2 M \left(\delta  e^{-\frac{r}{\lambda }}+1\right)}{(\delta +1) r},
\end{equation} is known as a Yukawa-like potential. Here $\delta,~~ \lambda$ represent the coefficients of Taylor expansion, $f'_0=1+\delta$, $\delta$ is a dimensionless parameter that characterizes the deviations from GR and governs the strength of the Yukawa-like correction, $\delta=0$ maps this metric to the Newtonian GR. Whereas $\lambda=[-f'_0/(6f''_0)]^{1/2}$, the parameter $\lambda$ denotes a characteristic length scale that controls the exponential decay of the modification. In the limit $\delta=0$, the potential reduces to the Newtonian potential, consistent with GR. The roots of the event horizon are given in Ref.~\cite{yukawaMetric}. They are found by solving \( g_{tt}(r) = 0 \), which leads to a transcendental equation due to the nonlinear nature of the Yukawa potential. Figure~\ref{fig:ricci} shows the behavior of the Ricci scalar for different values of $\delta$. The curvature increases near the horizon and decreases asymptotically with $r$. 

\section{Test particle dynamics around an ENLMY Black Hole }
\subsection{Equation of Motion}
The static and spherically symmetric BH metric admits useful symmetries, namely, the time translation and spatial rotation about the symmetry axis. The associated conserved quantities can be determined using the Killing vectors \cite{hobson2006general}, given respectively
\begin{equation}
\xi_{(t)}^{\mu}\partial_{\mu}=\partial_{t} , \quad
\xi_{(\phi)}^{\mu}\partial_{\mu}=\partial_{\phi},
\end{equation}
where
$\xi_{(t)}^{\mu}=(1,~0,~0,~0)$ and $\xi_{(\phi)}^{\mu}=(0,~0,~0,~1)$.
These symmetries give rise to conserved quantities: the specific energy ${\cal E}\equiv - p_\mu \xi_{(t)}^{\mu}/m $ and the specific azimuthal angular momentum $ {\cal L}_z={\cal L}\equiv p_\mu \xi_{(\phi)}^{\mu}/m$ of the moving particle. Therefore, we have
\begin{align}\label{dot}
\dot{t}=\frac{\cal{E}}{1+\Phi(r)}, \quad \dot{\phi}=\frac{{\cal{L}}}{r^2 \sin^2\theta}.
\end{align}

Here, an overdot denotes differentiation with respect to the proper time coordinate
$\tau$. To further investigate the geodesic motion, we adopt the Hamiltonian formalism $\mathcal{H}=\frac{1}{2}g^{\mu\nu}p_\mu p_\nu$ \cite{goldstein2011classical}; therefore, the Hamiltonian corresponding to our metric is
\begin{align}\label{5}
   \mathcal{H}=\frac{1}{2}\left[-\frac{{p_t}^2}{[1+\Phi(r)]}+\frac{{p_r}^2}{[1-\Phi(r)]} +\frac{{p_\theta}^2}{r^2}+\frac{{p_\phi}^2}{r^2 \sin^2\theta}\right],
\end{align}
and remains conserved along the particle’s trajectory. In this framework, the cyclic nature of $t$ and $\phi$ confirms the conservation of the conjugate momenta $p_t=-{\cal E}$ and $p_\phi={\cal L}$, consistent with the Killing vector analysis. Using Hamilton’s equations $\dot{x}^\mu=\frac{\partial\mathcal{H}}{\partial p_\mu}$ and $\dot{p}_\mu=-\frac{\partial \mathcal{H}}{\partial x^\mu}$ \cite{goldstein2011classical}, we obtain
  \begin{align}\label{6}
      \dot{p}_t=\dot{p}_\phi=0,\quad \dot{r}=\frac{p_r}{[1-\Phi(r)]},
 \quad \dot{\theta}=\frac{p_\theta}{r^2}.   
 \end{align}
 By the normalization of the four-momentum, the Hamiltonian takes the fixed value  $\mathcal{H}=\frac{1}{2}g^{\mu\nu}p_\mu p_\nu=\frac{1}{2}\epsilon$, with 
$\epsilon=-1$ for massive and $\epsilon=0$ for massless particles. Therefore, we have
\begin{align}
\label{7}
    \dot{p}_r &=\frac{1}{2}\left[-\frac{p_{r}^2}{[1-\Phi(r)]^\prime}+\frac{2p_\theta^2}{r^3}-\frac{2\Theta}{r^3}+\partial_r\left(\frac{\mathcal{
    R}}{[1+\Phi(r)]}\right)\right],\\
    \dot{p}_\theta &=\frac{1}{2r^2}\partial_\theta\left(\Theta\right),
\end{align}
where $``\prime"$ denotes the differentiation with respect to the radial coordinate $r$ and
\begin{align}\label{8}
    \mathcal{R}(r) &={\cal E}^2-[1+\Phi(r)]\left(1+\frac{{\cal L}^2}{r^2}+\frac{\mathcal{K}}{r^2}\right),\\
    \Theta(\theta) &=\mathcal{K}-\frac{{\cal L}^2
    \cos^2\theta}{\sin^2\theta}.
    \end{align}
 Here, $\mathcal{K}$ is the separation constant. From now onward, the motion of the test particles at the equatorial plane will be considered by a constant plane, choosing $\theta=\pi/2$ and $\mathcal{K}=0$.

\section{Effective Potential and Circular Orbits}

\begin{figure}[H]
  \centering
\includegraphics[width=0.68\linewidth]{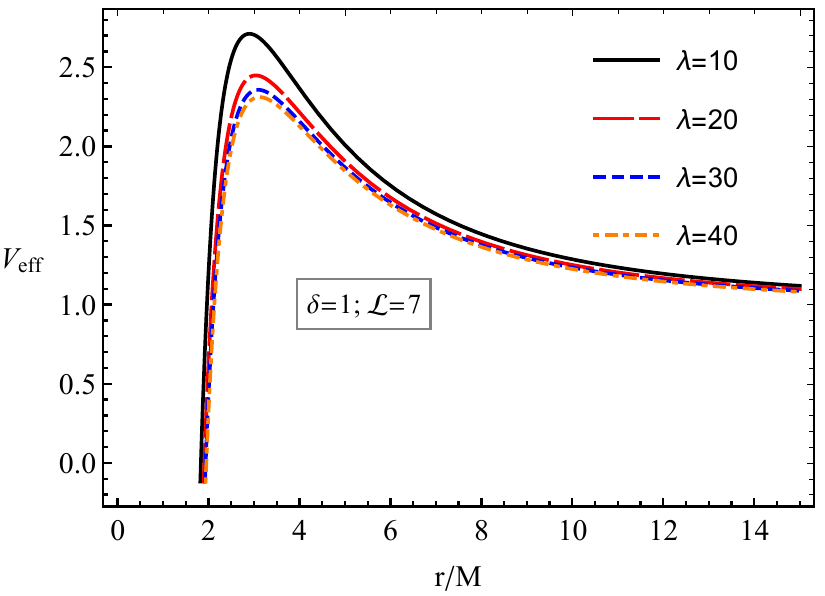} \includegraphics[width=0.68\linewidth]{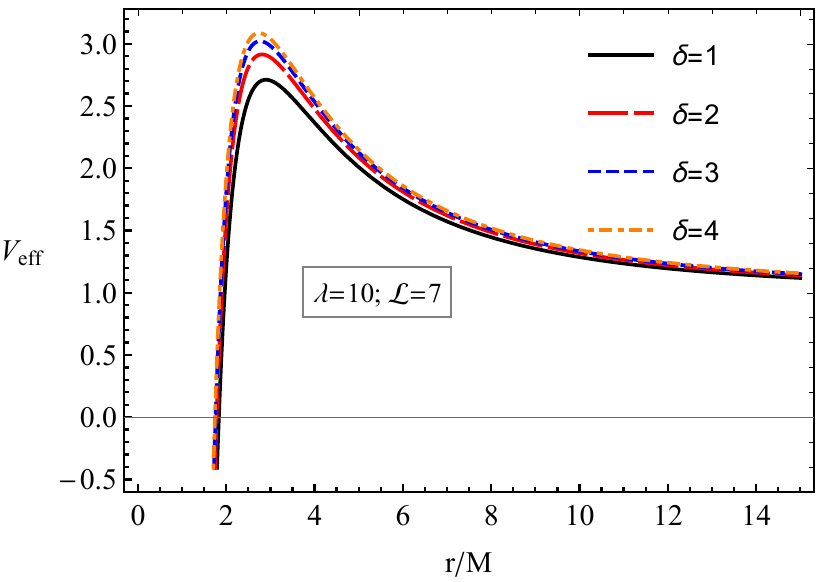}
\caption{Effective potential for radial motion of test particles around the ENLMY BH for various values of parameter $\lambda$ and $\delta$.}  \label{effective}
\end{figure}

 By applying the Hamiltonian constraint $g^{\mu\nu}p_{\mu}p_\nu=\epsilon$, the particle's trajectory can be analyzed using an effective potential $V_{\text{eff}}$ through a corresponding first-order equation \cite{babar2017periodic}. Therefore, we have \cite{yukawa3}
\begin{eqnarray}\label{4}
\mathcal{E}^{2}={{\dot{r}}^2}{[1-\Phi^2(r)]}+ V_\text{eff};\quad
V_\text{eff}= {{[1+\Phi(r)]}}\left(1 +\frac{{\cal L}^2}{r^2}\right).
\end{eqnarray}
 In Fig.~\ref{effective}, we analyze the effective potential of the radial motion of the test particles around the ENLMY BH for different values of the BH charge and the parameters $\lambda$ and $\delta$. As expected for an asymptotically flat spacetime, it is evident that $V_{\text{eff}}\rightarrow1$ as $r\rightarrow\infty$. Consequently, particles with energy ${\cal E}>1$ can escape to infinity, while those with ${\cal E}<1$ remain gravitationally bound. Therefore, ${\cal E}=1$ serves as the critical energy that separates bound and unbound orbits.
 \par
\begin{figure}[H]
  \centering
\includegraphics[width=0.68\linewidth]{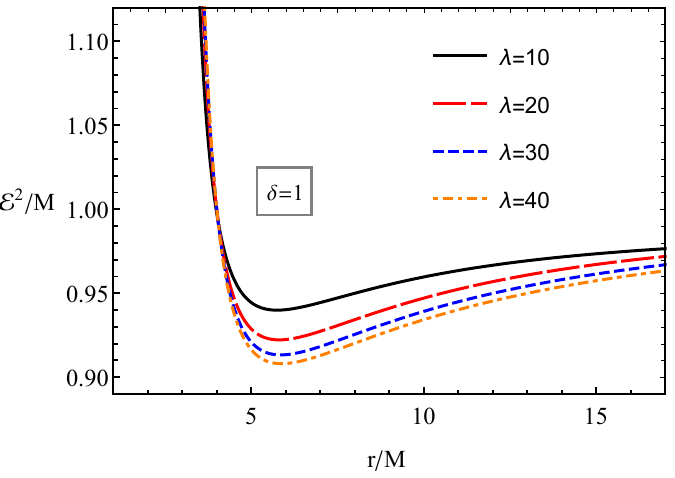} \includegraphics[width=0.68\linewidth]{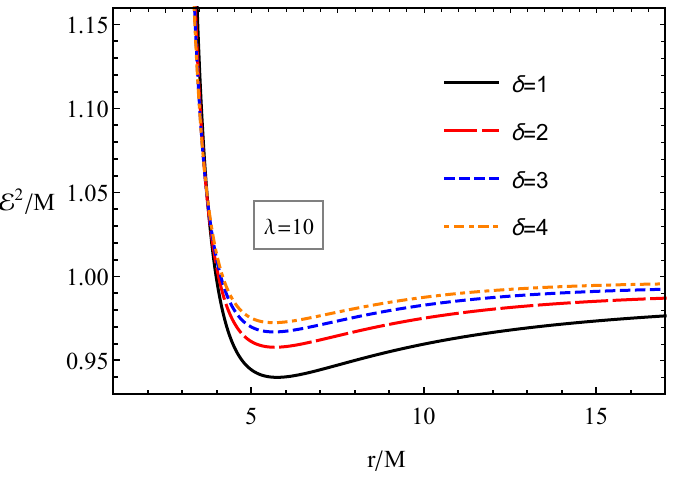}
	\caption{The radial dependence of the specific energy of a particle in circular orbits around the ENLMY BH for different values of $\lambda$ and $\delta$.} \label{energy}
\end{figure}

 The motion of test particles in circular orbits can be studied by solving the equation $\partial_rV_{\text{eff}}=0$, which gives the critical values of the angular momentum of the particles in the circular motion ${\cal L}_{\text{cr}}={\cal L}$. To analyze circular motion, we consider the conditions under which the radial velocity vanishes, i.e., $\dot{r}=0$, and there are no radial accelerations, $\ddot{r}=0$. These conditions yield the radial profiles of the specific angular momentum and specific energy for particles in circular orbits confined to the equatorial plane ($\theta=\frac{\pi}{2}$), expressed as follows
\begin{align}\label{L & E} 
\frac{{\cal L}^2}{M} &= -\frac{M r^2 \left[\lambda  \left(\delta +e^{r/\lambda }\right)+\delta  r\right]}{\delta  M (3 \lambda +r)-\lambda e^{r/\lambda } (-3 M+\delta  r+r)},\\
\frac{{\cal E}^2}{M} &= \frac{\lambda  e^{-\frac{r}{\lambda }} \left[e^{r/\lambda } (-2 M+\delta  r+r)-2 \delta  M\right]^2}{(\delta +1) r \left[\lambda  e^{r/\lambda } (-3 M+\delta  r+r)-\delta  M (3 \lambda +r)\right]}.
\end{align}
Fig.~\ref{energy} illustrates the radial dependence of the specific energy of particles in circular orbits around an ENLMY BH, showing how the energy profile varies with different values of $\lambda$ and $\delta$. It shows that increasing the Yukawa parameters $\lambda$ and $\delta$ leads to a higher specific energy and shifts the location of the minimum energy (stable orbit) to larger radii.

\begin{figure}[H]
   \centering
\includegraphics[width=0.68\linewidth]{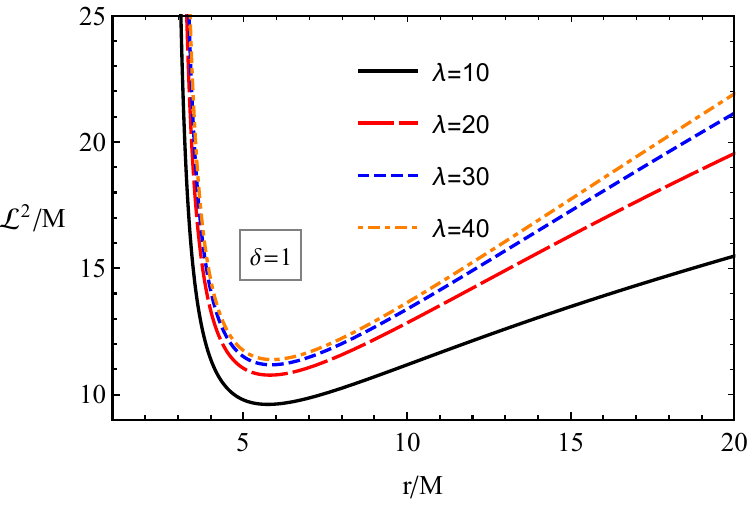} \includegraphics[width=0.68\linewidth]{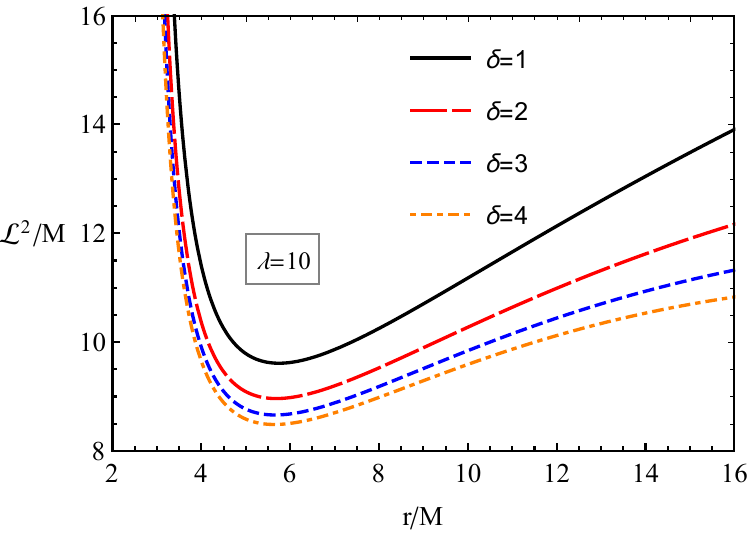}
	\caption{The radial dependence of the specific angular momentum of a particle in circular orbits around the ENLMY BH for different values of $\lambda$ and $\delta$.}\label{angular}
\end{figure}

In Fig.~\ref{angular}, we display the specific angular momentum of the particle corresponding to the circular motion as a function of the radial coordinate for different values of $\lambda$ and $\delta$. As the Yukawa parameters $\lambda$ and $\delta$ increase, the minimum of the angular momentum shifts to larger radii.
\subsection{Innermost Stable and Innermost Bound Circular Orbits}
To determine the radius of the ISCO, we begin with the radial equation of motion $\dot{r}^2 ={\cal E}^2 - V_{\text{eff}}(r)$. The ISCO defines the minimum radius at which a test particle can maintain a stable circular orbit around a massive object and is characterized by the following conditions \cite{bambi2017black,gao2020bound,Zahra:2025fvq}
\begin{align}\label{14}
\frac{dV_{\text{eff}}}{dr} = 0, \quad \frac{d^2V_{\text{eff}}}{dr^2} = 0, \quad V_{\text{eff}} ={\cal E}^2.
\end{align}
Applying these conditions leads to the following equation for the ISCO radius
\begin{align}\label{15}
& -2 \delta^2 M \left(3 \lambda^2 + r^2 + 5 \lambda r\right) 
- \delta e^{r/\lambda} \bigg[(\delta + 1) r \left(-\lambda^2 + r^2 - \lambda r\right) \nonumber\\
 \quad &- 2M (r - 6\lambda)(\lambda + r)\bigg]
+ \lambda^2 e^{2r/\lambda} (-6M + \delta r + r) = 0.
\end{align}
\indent The IBCO represents the smallest unstable circular orbit with finite energy, specifically ${\cal E}=1$. At this radius, a test particle has the same energy as it would have at infinity, indicating no energy loss during infall. Thus, the IBCO marks the threshold between bound and plunging orbits. For IBCO, the conditions for particle motion are \cite{bambi2017black,gao2020bound,shabbir2025periodic}
    \begin{align}\label{16}
    \frac{dV_{\text{eff}}}{dr}=0, \quad V_{\text{eff}}={\cal E}^2=1.
    \end{align}
Applying these conditions leads to the following equation for the IBCO radius
\begin{align}\label{17}
&\lambda  e^{-\frac{r}{\lambda }} \left[e^{r/\lambda } (-2 M+\delta  r+r)-2 \delta  M\right]^2-(\delta +1) r  \nonumber\\
\quad &\times \left[\lambda  e^{r/\lambda } (-3 M+\delta  r+r)-\delta  M (3 \lambda +r)\right]=0.
\end{align}
\begin{table*}
\centering
\caption{Values of $r_{\text{ISCO}}/M$, $r_{\text{IBCO}}/M$, ${\cal L}_{\text{ISCO}}/M$, and ${\cal L}_{\text{IBCO}}/M$ for different values of $\delta$, with fixed $\lambda=10$ and $\lambda=60$. Similar results are obtained for other values of $\lambda$.}
\label{tab:1}
\begin{tabular}{c|cccc|cccc}
\toprule
\multirow{2}{*}{$\delta$} & \multicolumn{4}{c|}{$\lambda=10$} & \multicolumn{4}{c}{$\lambda=60$} \\
\cline{2-9}
 & $r_{\text{ISCO}}$ & $r_{\text{IBCO}}$ & ${\cal L}_{\text{ISCO}}$ & ${\cal L}_{\text{IBCO}}$ & $r_{\text{ISCO}}$ & $r_{\text{IBCO}}$ & ${\cal L}_{\text{ISCO}}$ & ${\cal L}_{\text{IBCO}}$ \\
\midrule
-0.9 & 54.9635 & 35.6635 & 34.1538 & 39.1526 & 8.05754 & 4.87847 & 5.06249 & 7.41828 \\
-0.8 & 14.2149 & 9.91364 & 12.4269 & 15.4844 & 6.77431 & 4.21682 & 4.02719 & 5.19746 \\
-0.7 & 8.77196 & 5.96022 & 6.77339 & 8.83245 & 6.43011 & 4.08847 & 3.77128 & 4.65352 \\
-0.6 & 7.39147 & 4.92928 & 5.16396 & 6.60944 & 6.27008 & 4.04298 & 3.65537 & 4.40733 \\
-0.5 & 6.80445 & 4.50169 & 4.46530 & 5.57499 & 6.17758 & 4.02254 & 3.58926 & 4.26678 \\
-0.4 & 6.48776 & 4.28236 & 4.08171 & 4.98607 & 6.11732 & 4.01211 & 3.54652 & 4.17582 \\
-0.3 & 6.29269 & 4.15655 & 3.84063 & 4.60733 & 6.07493 & 4.00637 & 3.51663 & 4.11213 \\
-0.2 & 6.16193 & 4.07968 & 3.67545 & 4.34351 & 6.04350 & 4.00307 & 3.49455 & 4.06502 \\
-0.1 & 6.06899 & 4.03109 & 3.55533 & 4.14918 & 6.01926 & 4.00113 & 3.47756 & 4.02877 \\
 0.0 & 6.00000 & 4.00000 & 3.46410 & 4.00000 & 6.00000 & 4.00000 & 3.46410 & 4.00000 \\
 0.1 & 5.94705 & 3.98031 & 3.39248 & 3.88182 & 5.98432 & 3.99936 & 3.45317 & 3.97662 \\
 0.2 & 5.90533 & 3.96831 & 3.33476 & 3.78582 & 5.97132 & 3.99902 & 3.44411 & 3.95723 \\
 0.3 & 5.87173 & 3.96167 & 3.28727 & 3.70627 & 5.96036 & 3.99889 & 3.43648 & 3.94090 \\
 0.4 & 5.84418 & 3.95883 & 3.24751 & 3.63924 & 5.95090 & 3.99889 & 3.42997 & 3.92696 \\
 0.5 & 5.82125 & 3.95874 & 3.21373 & 3.58196 & 5.94289 & 3.99897 & 3.42434 & 3.91491 \\
 0.6 & 5.80191 & 3.96065 & 3.18469 & 3.53244 & 5.93582 & 3.99911 & 3.41944 & 3.90440 \\
 0.7 & 5.78541 & 3.96405 & 3.15944 & 3.48919 & 5.92960 & 3.99928 & 3.41513 & 3.89515 \\
 0.8 & 5.77120 & 3.96854 & 3.13730 & 3.45107 & 5.92408 & 3.99948 & 3.41130 & 3.88694 \\
 0.9 & 5.75886 & 3.97384 & 3.11720 & 3.41721 & 5.91914 & 3.99969 & 3.40788 & 3.87961 \\
 \bottomrule
\end{tabular}
\end{table*}
\begin{figure}[H]
   \centering
\includegraphics[width=0.68\linewidth]{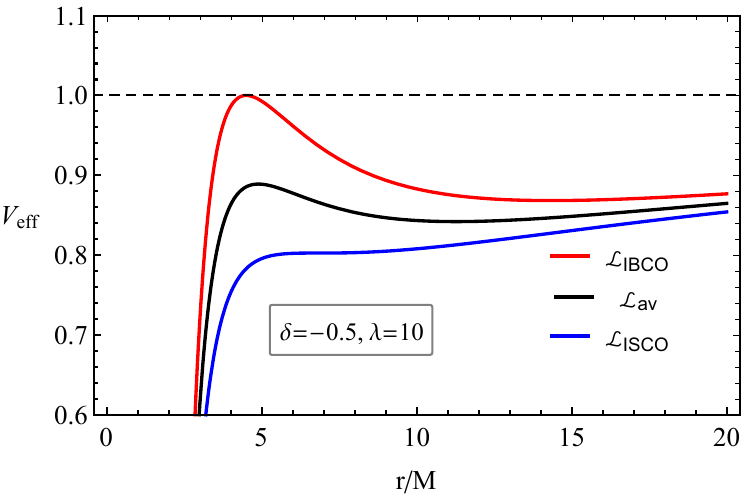} \includegraphics[width=0.68\linewidth]{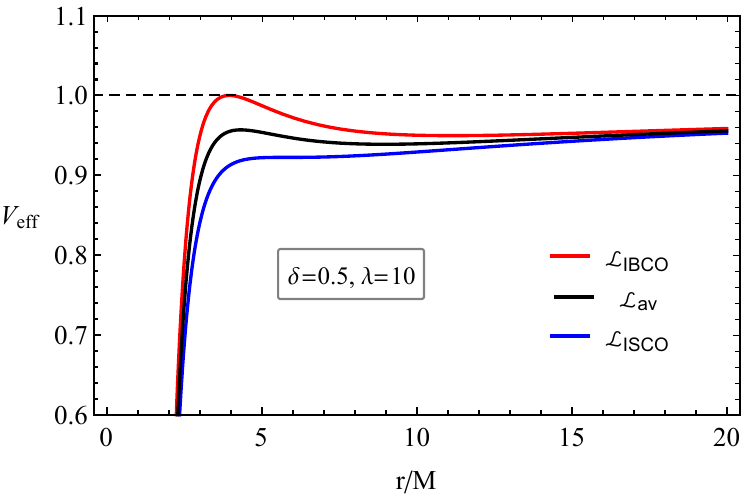}
	\caption{Behavior of the effective potential for various angular momenta, plotted for $\lambda = 10$ and $\delta = -0.5$ (top), $ 0.5$ (bottom).}
\label{ang}
\end{figure}
The transcendental Eqs. (\ref{15}) and (\ref{17}) generally do not admit analytical solutions. However, for fixed values of the parameters $\lambda$ and $\delta$, the radii corresponding to the ISCO and IBCO can be determined numerically. To reduce the number of free parameters, we define a new angular momentum given as \cite{levin2008periodic,shabbir2025periodic}
\begin{align}\label{18}  
{\cal L}_{\text{av}}=\frac{{\cal L}_{\text{ISCO}}+{\cal L}_{\text{IBCO}}}{2}. 
\end{align}
For each chosen pair of $\lambda$ and $\delta$, one can uniquely determine ${\cal L}_{\text{ISCO}}$ and ${\cal L}_{\text{IBCO}}$, such that the angular momentum of the test particles satisfies the condition ${\cal L}_{\text{ISCO}}<{\cal L}_{\text{av}}<{\cal L}_{\text{IBCO}}$ \cite{levin2008periodic,shabbir2025periodic}.
In Table~\ref{tab:1}, we present the values of the ISCO radius, the IBCO radius, ${\cal L}_{\text{ISCO}}$, and ${\cal L}_{\text{IBCO}}$ for various parameter sets. In Fig.~\ref{ang}, the effective potential is shown for different values of angular momentum. For ${\cal L}_{\text{IBCO}}$, the maximum of the potential corresponds to $V_{\text{eff}}=1$ for all values of $\lambda$ and $\delta$.
\section{Periodic Orbits}

 In this section, we investigate the periodic orbits of test particles around the ENLMY BH, employing the classification scheme proposed by Levin \cite{levin2008periodic}. Bound orbits are characterized by oscillations in both the radial coordinate $r$ and the angular coordinate $\phi$. A periodic orbit arises when the ratio of the angular to radial orbital frequencies, given by $q=\frac{\omega_\phi}{\omega_r}$, is a positive rational number. Under this condition, the particle's trajectory becomes closed, allowing it to return to its initial position after a finite interval, thereby repeating its motion. The rational number $q$ can be expressed in terms of three integers $(z,w,v)$ given by
 \begin{equation}\label{19}
q = w + \frac{v}{z} = \frac{\Delta\phi}{2\pi} - 1,
\end{equation}
where
\begin{equation}\label{20}
\Delta\phi = 2 \int_{r_-}^{r_+} \frac{\dot{\phi}}{\dot{r}}\,dr 
= 2 \int_{r_-}^{r_+} \frac{{\cal L}}{r^2 \sqrt{\frac{1}{1-\Phi^2(r)}\left[{\cal E}^2 - \left[1 + \Phi(r)\right] \left(1 + \frac{{\cal L}^2}{r^2} \right)\right] }}dr.
\end{equation}
The triplet of integers $(z,w,v)$ provides a geometric interpretation of the orbital structure: $z$ denotes the number of ``zooms", $w$ indicates the number of ``whirls", and $v$ specifies the number of distinct ``vertices". The radial motion is bounded between two turning points $r_+$ and $r_-$, which lie between the ISCO and the IBCO. These turning points can be found by solving the condition $\dot{r}^2=0$, where $r_{\pm}$ are the roots of the equation ${\cal E}^2=V_{\text{eff}}$. From Eqs.~(\ref{19}) and (\ref{20}), the rational number $q$ becomes
\begin{equation}
\label{eq21}
q = \frac{1}{\pi} \int_{r_-}^{r_+} \frac{{\cal L}}{r^2 \sqrt{\frac{1}{1-\Phi^2(r)}\left[{\cal E}^2 - \left[1 + \Phi(r) \right] \left(1 + \frac{{\cal L}^2}{r^2} \right)\right] }}\,dr - 1.
\end{equation}
From Eq.~(\ref{eq21}), the rational number $q$ depends on ${\cal E}$, ${\cal L}$, $M$, $\lambda$, and $\delta$. If we further consider ${\cal L}={\cal L}_{\text{av}}$, where ${\cal L}_{\text{av}}$ is a fixed value determined by the parameters $M$, $\lambda$, and $\delta$, then the rational number $q$ becomes dependent only on the orbital energy ${\cal E}$. Using Eq.~(\ref{eq21}), we construct Tables~\ref{tab:2} and \ref{tab:3} for various values of $\delta$ with $\lambda=10$ and $\lambda=60$, respectively.

\begin{figure*}
   \centering
\includegraphics[width=0.4\linewidth]{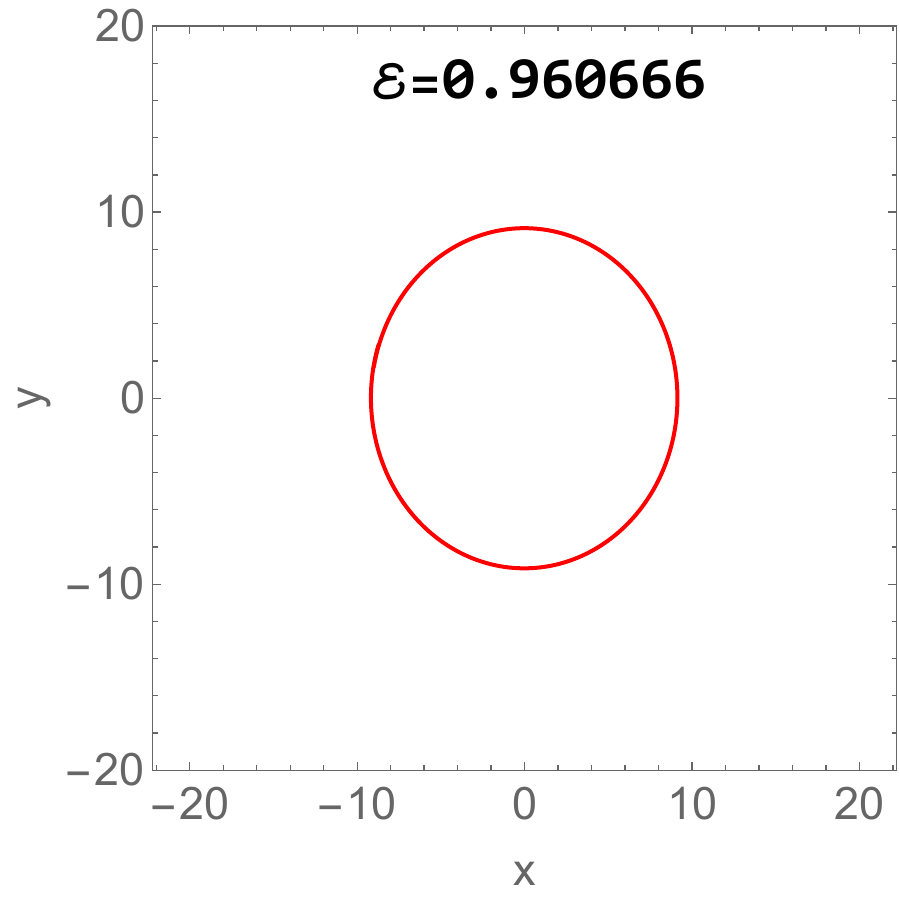}
\includegraphics[width=0.4\linewidth]{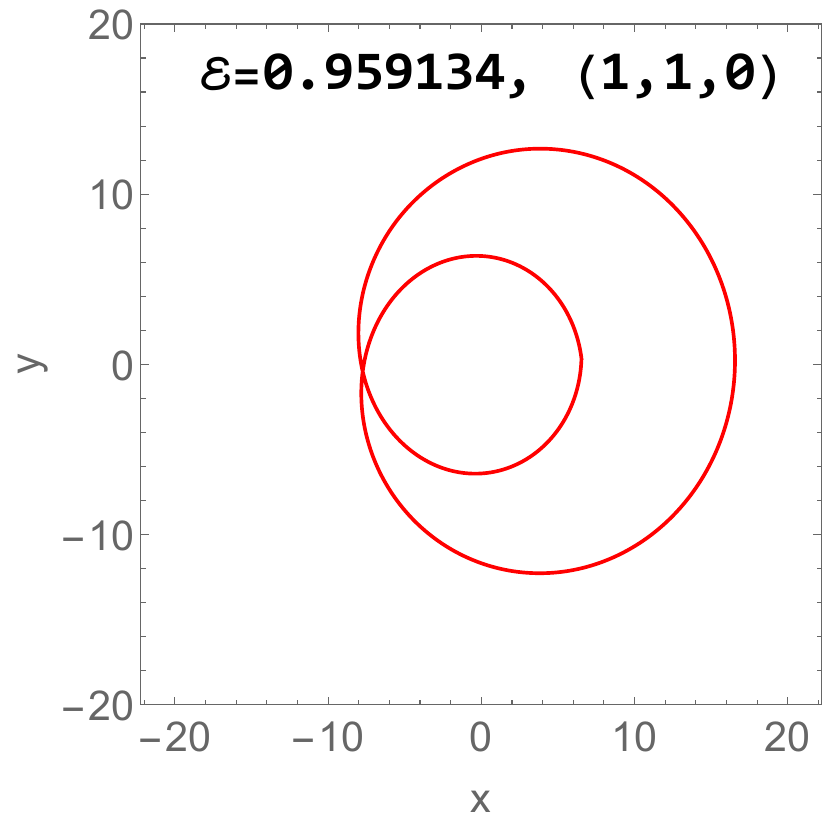}
/\includegraphics[width=0.4\linewidth]{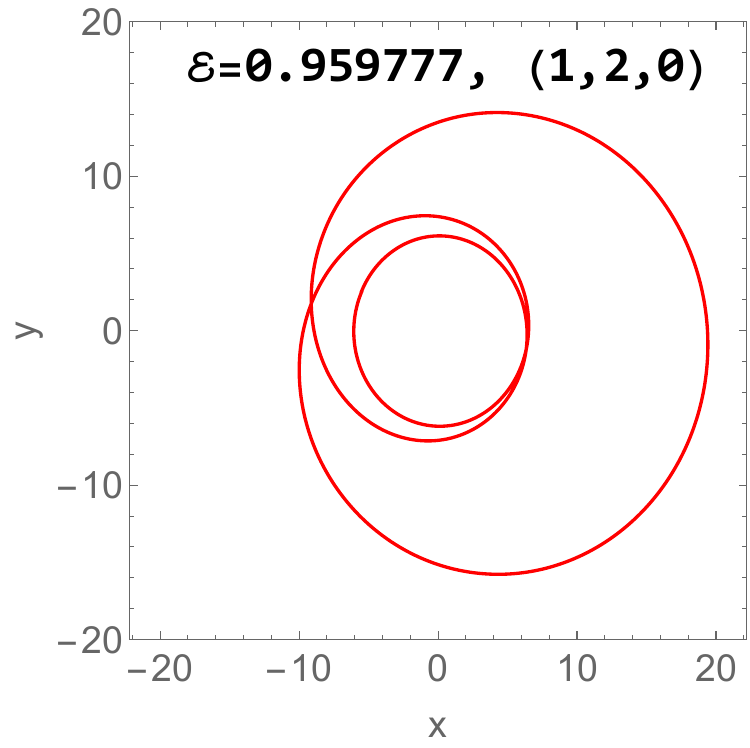}
\includegraphics[width=0.4\linewidth]{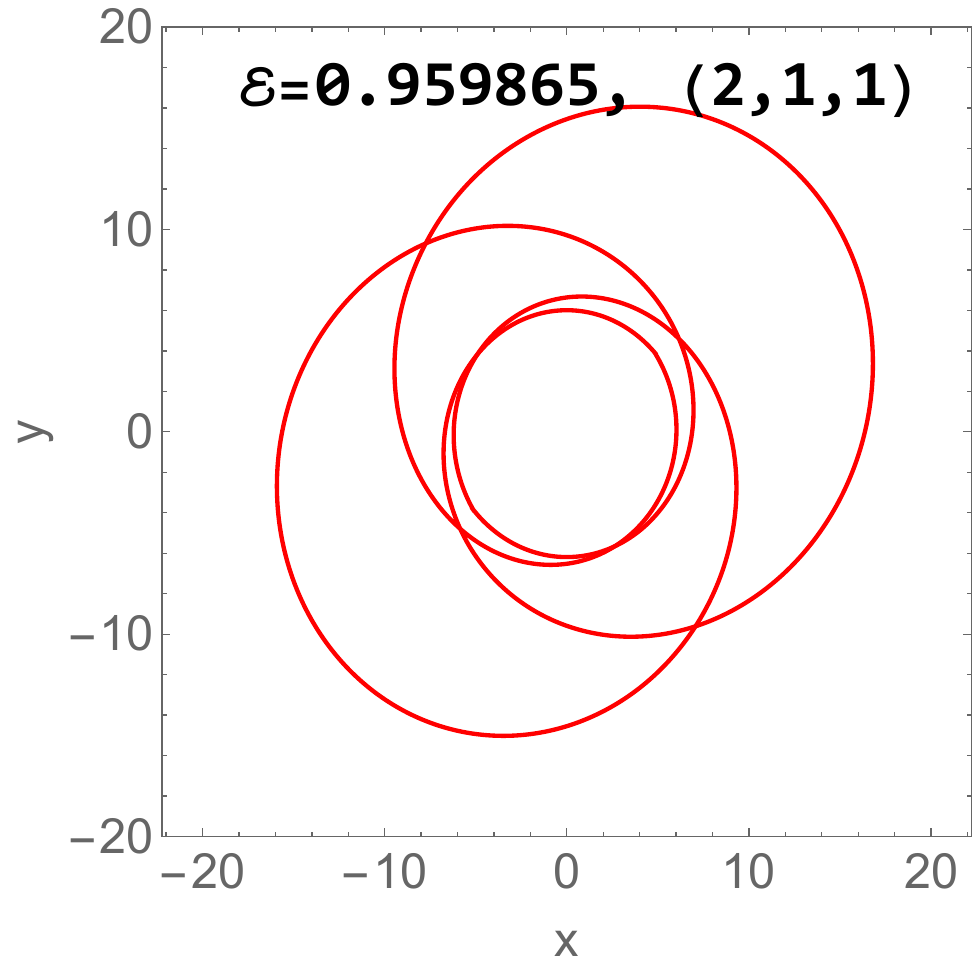}
\includegraphics[width=0.4\linewidth]{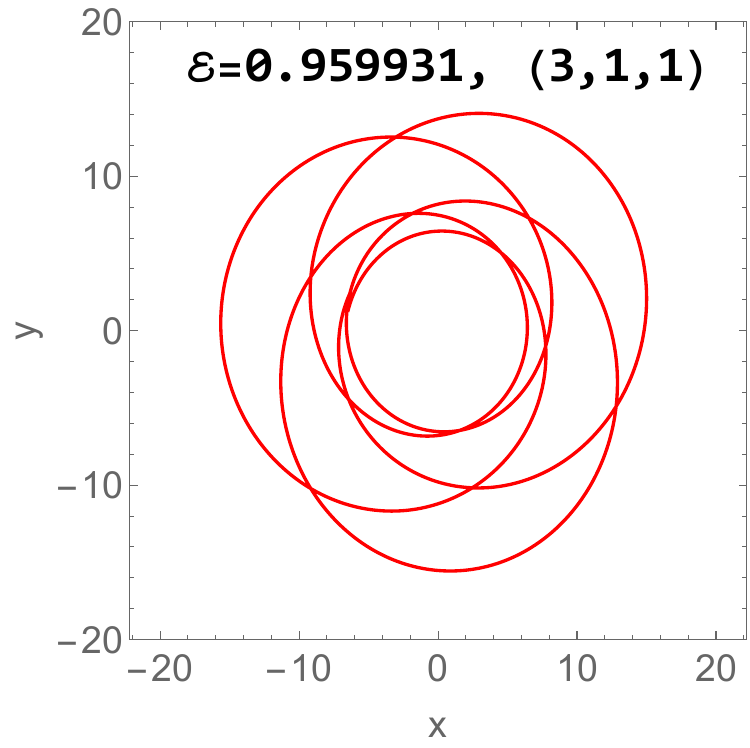}
\includegraphics[width=0.4\linewidth]{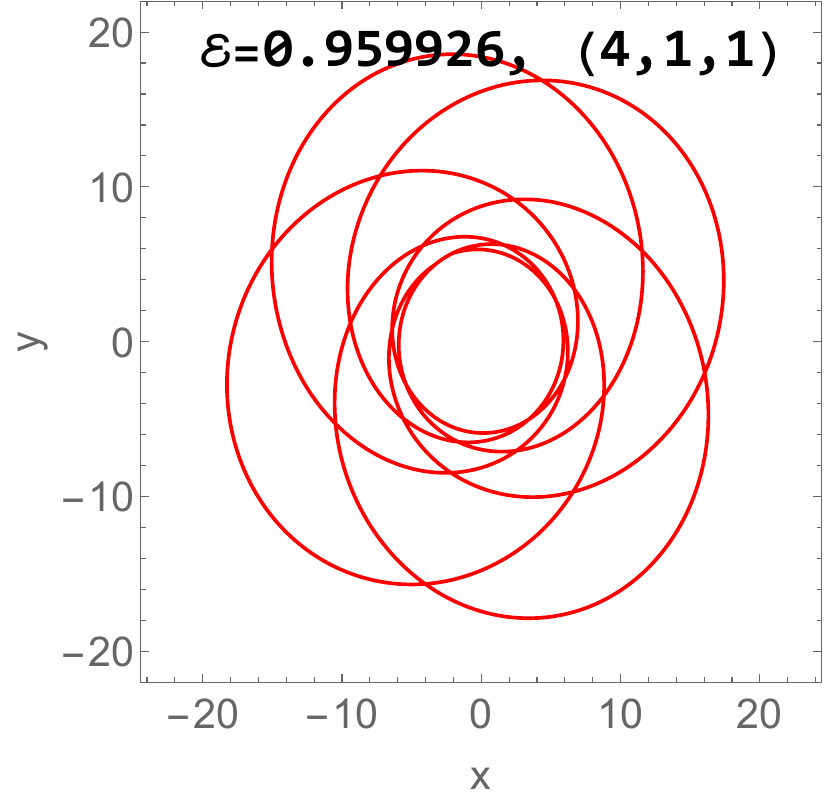}
\caption{ Zoom-whirl periodic orbits for $(z,w,v)$ with $z=1,2,3,4$, $w=1,2$, $v=1$ for $\lambda=60$ and $\delta=0.5$, plotted using $x=\frac{r}{M}\cos \phi, y=\frac{r}{M}\sin\phi$. } \label{PO}
\end{figure*}

From this table, it is evident that for small values of $\lambda$ (i.e., $\lambda=10$, as presented in the Table~\ref{tab:2}), the energy associated with periodic orbits around the ENLMY BH is higher than that of the corresponding orbits in the Schwarzschild spacetime, regardless of the sign of $\delta$. In contrast, for larger values of $\lambda$ (i.e., $\lambda=60$, as presented in the Table~\ref{tab:3}), the energy of periodic orbits in the ENLMY spacetime becomes lower than in the Schwarzschild case, for any value of $\delta$, whether positive or negative. This indicates that the stronger gravitational binding in the presence of Yukawa corrections highlights the influence of modified gravity on orbital dynamics.
Fig.~\ref{PO}  illustrates the periodic orbits of massive particles lying between the ISCO and IBCO around the ENLMY BH for 
$\lambda=60$ and $\delta=0.5$. Each orbit, identified by a triplet of integers, corresponds to a distinct energy value. After detailed analysis, we conclude that periodic orbits in ENLMY BH exhibit qualitative similarities to those in Schwarzschild and Kerr spacetimes. However, the presence of Yukawa corrections significantly alters the orbital energy profiles and enhances gravitational binding.
    \begin{table*}
\centering
    \caption{Energy values and corresponding angular momentum for periodic orbits with parameters $(z=1,2,3,w=1,v=1)$ around the ENLMY BH for $\lambda=10$ and various values of $\delta=0,0.1,0.2,0.3,0.4,0.5$.}
     \label{tab:2}
    \begin{tabular}{cccccc}
    \toprule
      $\delta$ & ${\cal L}_{\text{av}}$ &${\cal E}_{(1,1,0)}$  &${\cal E}_{(2,1,1)}$  & ${\cal E}_{(3,1,1)}$ & ${\cal E}_{(4,1,1)}$\\
      \midrule
       0& 3.732055 & 0.966348 &  0.968214  & 0.967969 & 0.967758 \\
      0.1 & 3.63715 & 0.970087 & 0.969362 & 0.969833     & 0.969833  \\
        0.2 & 3.56029 & 0.969985   & 0.969883 & 0.969405 & 0.969917  \\
         0.3& 3.49677 & 0.969986 &0.969974    & 0.96909   & 0.969707\\
         0.4& 3.443375 & 0.969182  & 0.969859  & 0.969893 & 0.968156  \\
         0.5& 3.397845 & 0.969953  & 0.969267 & 0.969956  & 0.969967 \\
         \bottomrule
    \end{tabular}
    \end{table*}
\begin{table*}
\centering
    \caption{Energy values and corresponding angular momentum for periodic orbits with parameters $(z=1,2,3,w=1,v=1)$ around the ENLMY BH for $\lambda=60$ and various values of $\delta=0,0.1,0.2,0.3,0.4,0.5$.}
     \label{tab:3}
    \begin{tabular}{cccccc}
    \toprule
      $\delta$ & ${\cal L}_{\text{av}}$ &${\cal E}_{(1,1,0)}$  &${\cal E}_{(2,1,1)}$  & ${\cal E}_{(3,1,1)}$ & ${\cal E}_{(4,1,1)}$\\
      \midrule
      0& 3.732055 & 0.966348  & 0.968214  & 0.967969 
 & 0.967758  \\ 
         0.1& 3.714895 & 0.959224  & 0.956707 & 0.959458
 & 0.957118 \\ 
        0.2 & 3.70067 & 0.958138  & 0.959965 & 0.959514  & 0.959925  \\
         0.3& 3.68869 & 0.959995 & 0.958101   & 0.95963  & 0.959914\\
         0.4& 3.678465 & 0.95996 & 0.958681  & 0.959698 & 0.959991 \\
         0.5& 3.669625 & 0.959134  & 0.959865  & 0.959931  & 0.959926 \\
         \bottomrule
    \end{tabular}
    \end{table*}

\section{Gravitational Wave Radiation from Periodic Orbits}
In this section, we investigate the GW radiation emitted by periodic orbits of a test particle in the background of the ENLMY BH spacetime. We consider an EMRI system, where the mass of the orbiting particle is negligible compared to that of the central BH (supermassive BH), allowing us to treat the particle as a perturbation to the ENLMY geometry. Under this approximation, when the energy ${\cal E}$ and angular momentum ${\cal L}$ of the particle vary slowly due to the emission of GW, the adiabatic approximation becomes valid. Through this approach, the slow drift in conserved quantities due to gravitational radiation can be traced, revealing how periodic orbits evolve. The transitions between these orbits correspond to the emission of gravitational radiation.

To compute the GWs emitted by periodic orbits around the ENLMY BH, we adopt the kludge waveform approach as described in \cite{babak2007kludge}. In this method, the smaller object is treated as a test particle moving in the fixed background of the BH. The periodic trajectories are first obtained by numerically solving the geodesic equations in the ENLMY spacetime. Once the orbital motion is determined, the GWs are computed using an approximate formula valid up to quadratic order \cite{maselli2022detecting,liang2023probing,tu2023periodic,yang2024gravitational}
\begin{align}
    h_{ij}=\frac{4\eta M}{D_L}\left(V_i V_j-\frac{m}{r}n_i n_j\right).
\end{align}
Here, $M$ denotes the mass of the central ENLMY BH, $m$ is the mass of the test particle, and $D_L$
  represents the luminosity distance to the EMRI system. The symmetric mass ratio is defined as $\eta=\frac{Mm}{\left(M+m\right)^2}$, while $V_i$ refers to the spatial velocity components of the test particle. The vector $n_i$ is the unit radial vector that points from the source to the observer, aligned with the direction of motion of the test particle. The resulting GW radiation can be projected onto a detector-adapted coordinate system to yield the two polarization modes of the waveform: the plus $(h_+)$ and cross $(h_\times)$ polarizations, given by \cite{maselli2022detecting,liang2023probing,tu2023periodic,yang2024gravitational}.
\begin{align}
    h_+ &=-\frac{2\eta}{D_L}\frac{M^2}{r}\left(1+\cos^2\iota\right)\cos\left(2\phi+2\zeta\right),\\
    h_\times &=-\frac{4\eta}{D_L}\frac{M^2}{r}\cos \iota\sin\left(2\phi+2\zeta\right),
\end{align}
where $\iota$ denotes the inclination angle between the orbital angular momentum of the EMRI system and the observer’s line of sight, while $\zeta$ represents the latitudinal viewing angle. To illustrate the GW waveforms produced by different periodic orbits and to investigate how the Yukawa-type corrections affect these signals, we consider an EMRI configuration in which the smaller body has mass $m\ll M$, with $M$ being the mass of the central ENLMY BH. For simplicity, we fix the inclination and latitudinal angles to $\iota,\zeta=\frac{\pi}{4}$. Moreover, the luminosity distance to the source is assumed to remain constant, with $D_L=200Mpc$. Under these assumptions, the resulting waveforms are \cite{shabbir2025periodic}
\begin{align}
\label{hplus}
h_+ \propto -\frac{\cos(2\phi+2\zeta)}{r},
\end{align}
\begin{figure}[H]
	\centering
	\begin{minipage}[b]{0.25\textwidth}
		\centering
		\includegraphics[width=\textwidth]{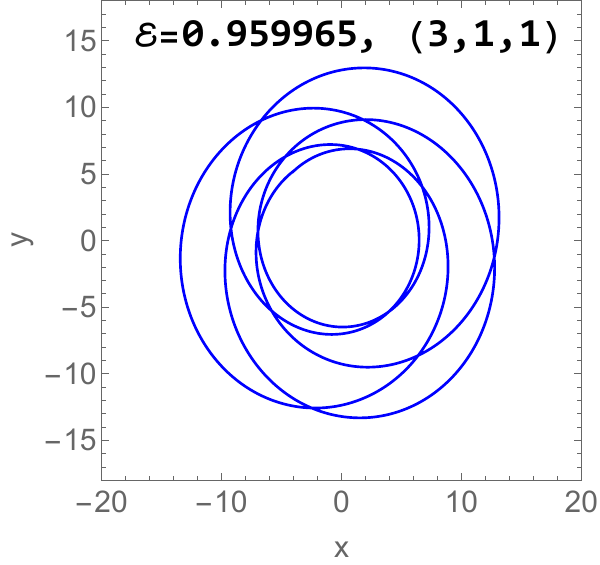}
	\end{minipage}
	\hfill
	\begin{minipage}[b]{0.3\textwidth}
		\centering
		\includegraphics[width=1\textwidth]{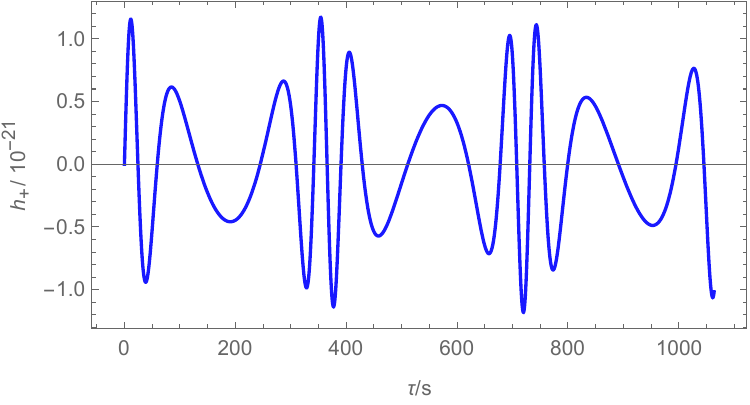}\\
		\includegraphics[width=1\textwidth]{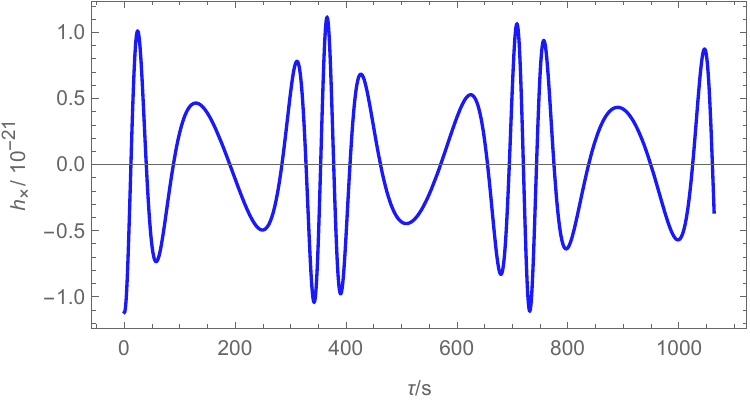}\\
	\end{minipage}
	\caption{The case $\delta=-0.05$, $\lambda=60$, ${\cal L}_{\text{av}}=3.742035$ and ${\cal E}=0.959965 $. The periodic orbit and the GW of the ENLMY BH for the triplet of integers $(3, 1, 1)$.}
	\label{Fig(-0.05)}
\end{figure}
\begin{figure}[H]
	\centering
	\begin{minipage}[b]{0.25\textwidth}
		\centering
		\includegraphics[width=\textwidth]{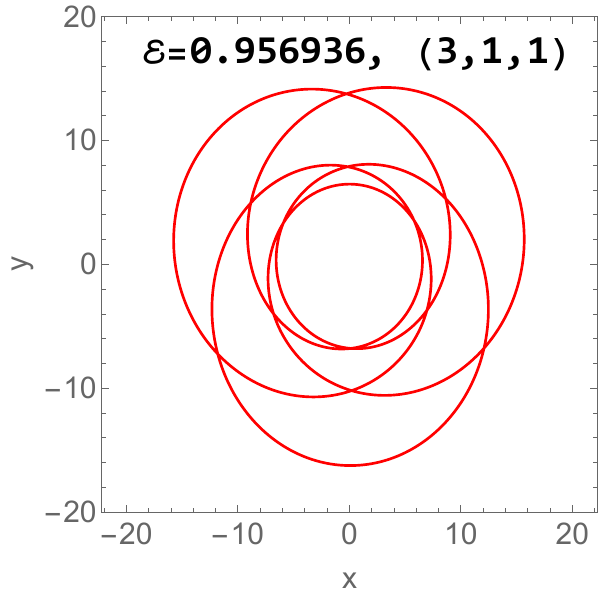}
	\end{minipage}
	\hfill
	\begin{minipage}[b]{0.3\textwidth}
		\centering
		\includegraphics[width=1\textwidth]{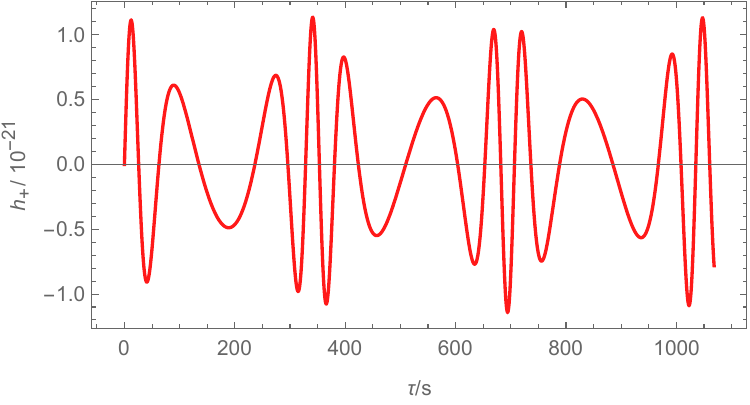}\\
		\includegraphics[width=1\textwidth]{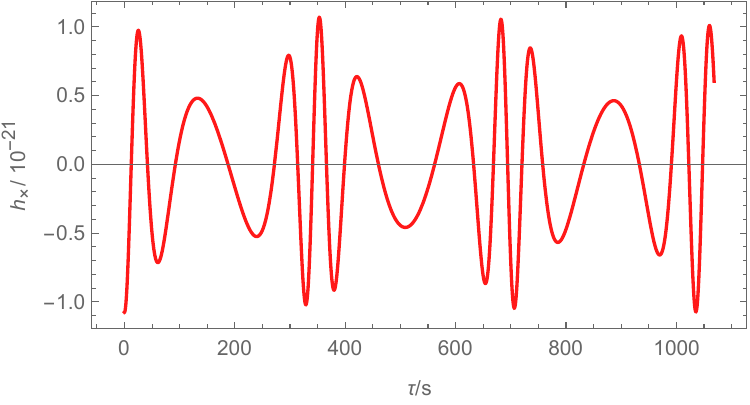}\\
	\end{minipage}
	\caption{The case $\delta=0.05$, $\lambda=60$, ${\cal L}_{\text{av}}=3.723045$ and ${\cal E}=0.956936 $. The periodic orbit and the GW of the ENLMY BH for the triplet of integers $(3, 1, 1)$.}
	\label{Fig(0.05)}
\end{figure}
and
\begin{align}\label{hcross}
h_\times \propto -\frac{\sin(2\phi+2\zeta)}{r}.
\end{align}
We restrict our analysis to plotting the right-hand sides of Eqs.~(\ref{hplus}) and (\ref{hcross}) as functions of the coordinate time $t$.

\begin{figure}[H]
	\centering
	\includegraphics[width=0.75\linewidth]{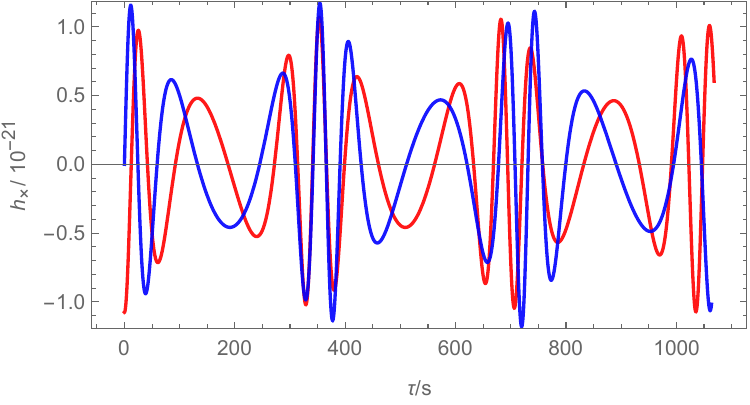}
	\includegraphics[width=0.75\linewidth]{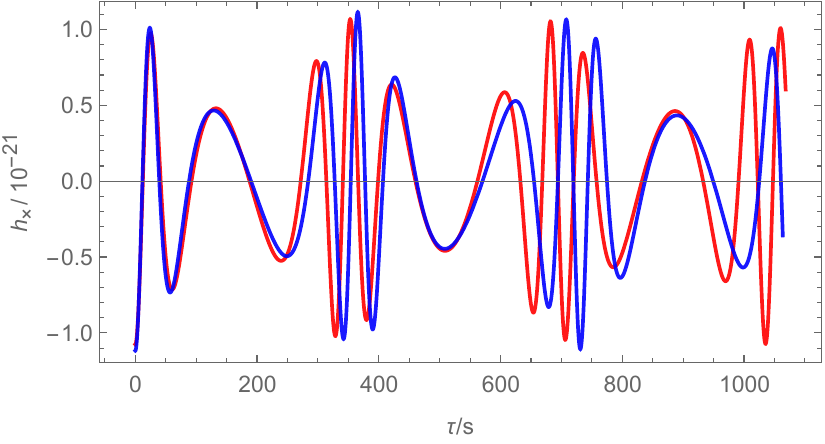}
	\caption{Upper Panel: Combination of the plots of $h_+$ in Figs.~\ref{Fig(-0.05)} and~\ref{Fig(0.05)}. Lower Panel: Combination of the plots of $h_\times$ in Figs.~\ref{Fig(-0.05)} and~\ref{Fig(0.05)}.}
	\label{Fig(311)}
\end{figure}
\begin{figure}[H]
	\centering
	\begin{minipage}[b]{0.25\textwidth}
		\centering
		\includegraphics[width=\textwidth]{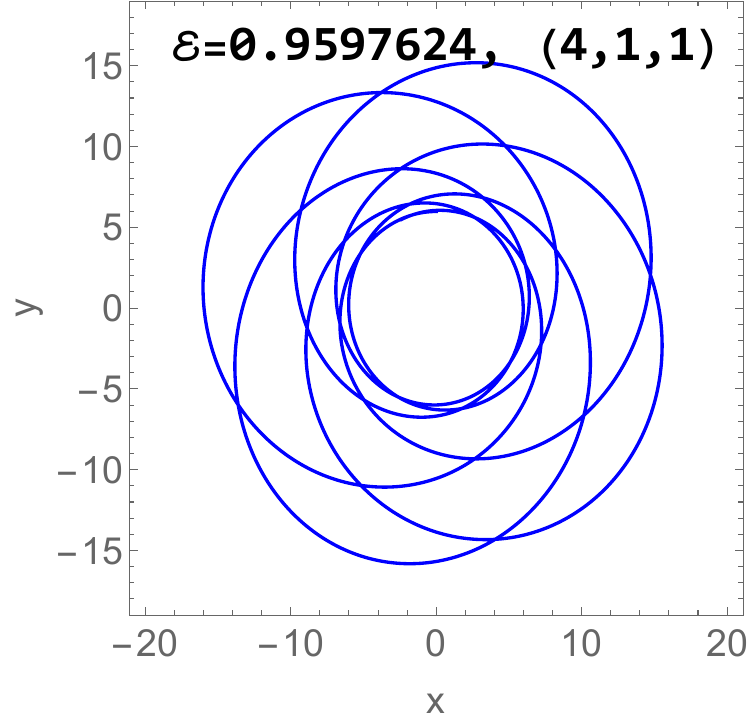}
	\end{minipage}
	\hfill
	\begin{minipage}[b]{0.3\textwidth}
		\centering
		\includegraphics[width=1\textwidth]{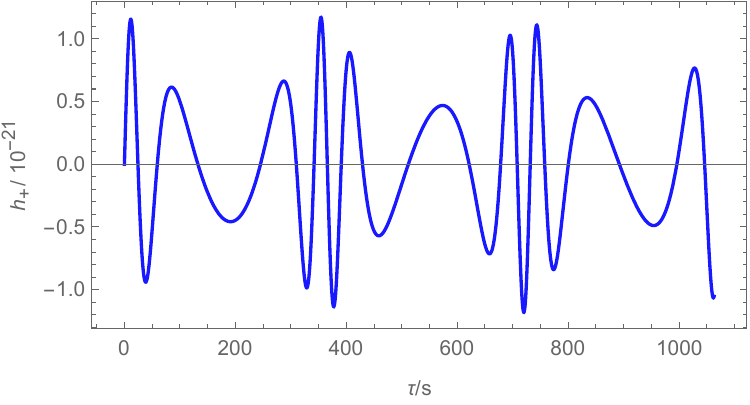}\\
		\includegraphics[width=1\textwidth]{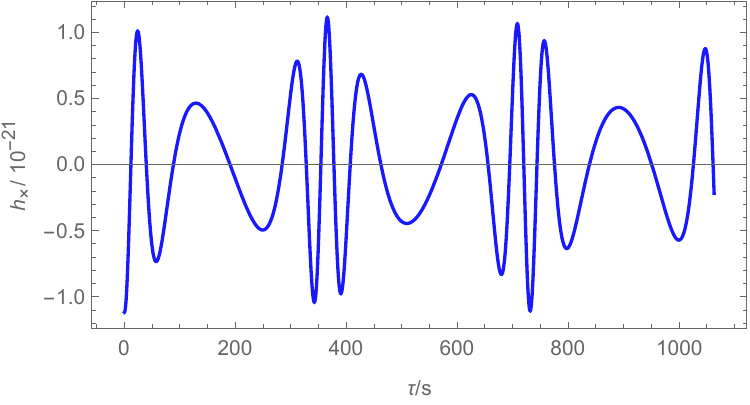}\\
	\end{minipage}
	\caption{The case $\delta=-0.05$, $\lambda=60$, ${\cal L}_{\text{av}}=3.742035$ and ${\cal E}=0.9597624 $. The periodic orbit and the GW of the ENLMY BH for the triplet of integers $(4, 1, 1)$.}
	\label{Fig(-0.05)411}
\end{figure}
\begin{figure}[H]
	\centering
	\begin{minipage}[b]{0.25\textwidth}
		\centering
		\includegraphics[width=\textwidth]{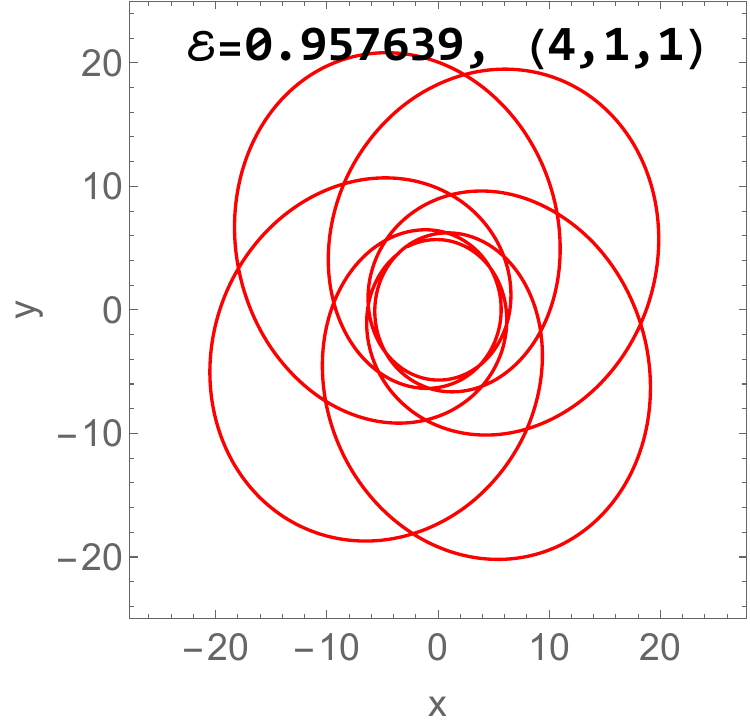}
	\end{minipage}
	\hfill
	\begin{minipage}[b]{0.3\textwidth}
		\centering
		\includegraphics[width=1\textwidth]{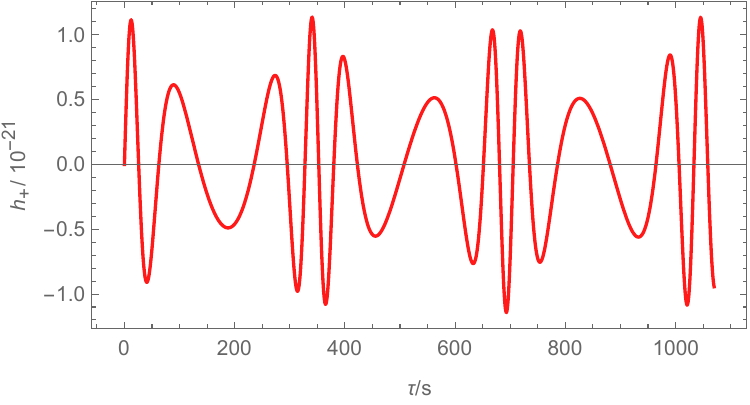}\\
		\includegraphics[width=1\textwidth]{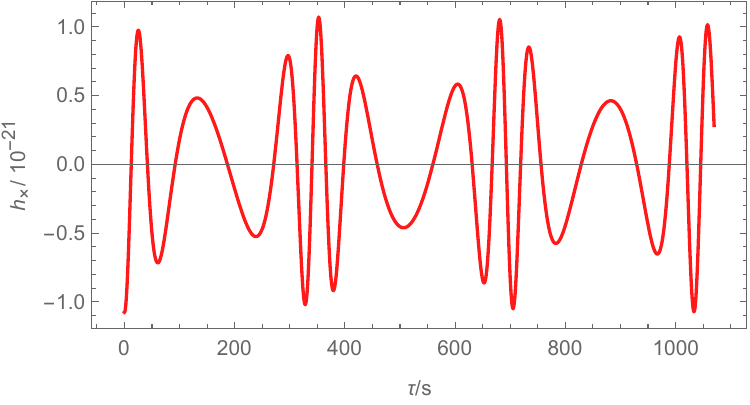}\\
	\end{minipage}
	\caption{The case $\delta=0.05$, $\lambda=60$, ${\cal L}_{\text{av}}=3.723045$ and ${\cal E}=0.957639 $. The periodic orbit and the GW of the ENLMY BH for the triplet of integers $(4, 1, 1)$.}
	\label{Fig(0.05)411}
    \end{figure}
 \begin{figure}[H]
	\centering
	\includegraphics[width=0.75\linewidth]{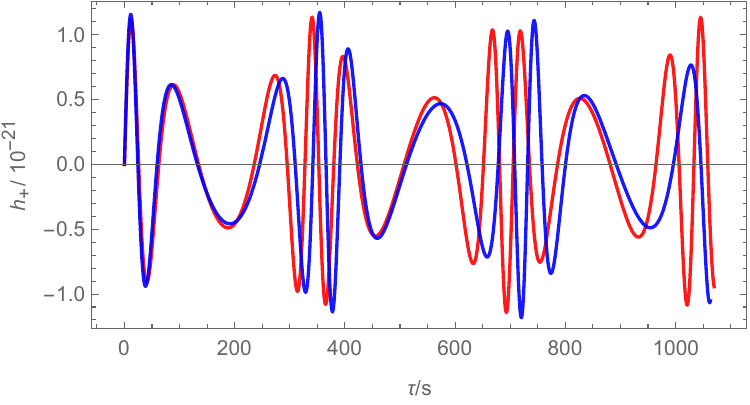}
	\includegraphics[width=0.75\linewidth]{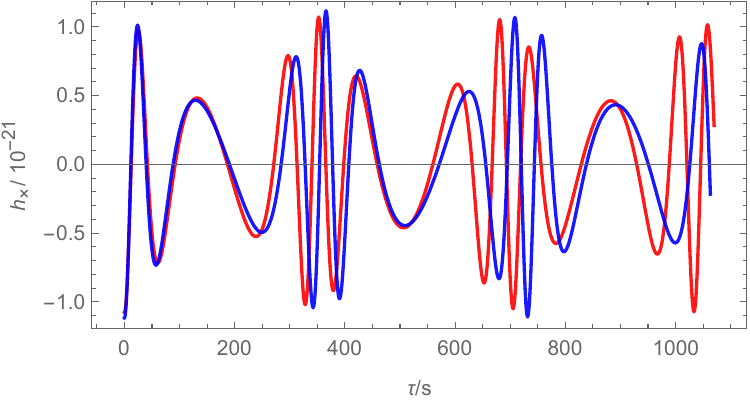}
	\caption{Upper Panel: Combination of the plots of $h_+$ in Figs.~\ref{Fig(-0.05)411} and~\ref{Fig(0.05)411}. Lower Panel: Combination of the plots of $h_\times$ in Figs.~\ref{Fig(-0.05)411} and~\ref{Fig(0.05)411}.}
	\label{Fig(411)}
\end{figure}
In Figs.~\ref{Fig(-0.05)}, \ref{Fig(0.05)} , and \ref{Fig(311)}, we consider small values of $|\delta|=0.05$, focusing on periodic orbits and the corresponding GWs associated with the triplet of integers $(3,1,1)$. In Figs.~\ref{Fig(-0.05)411}, \ref{Fig(0.05)411} , and \ref{Fig(411)}, we consider small values of $|\delta|=0.05$, focusing on periodic orbits and the corresponding GWs associated with the triplet of integers $(4,1,1)$.

In Figs.~\ref{Fig(-0.05)}-\ref{Fig(411)}, the periodic orbits, defined by a triplet of integers $(z,w,v)$, exhibit distinct zoom and whirl phases within each complete cycle. During the zoom phase, the particle follows an extended, elliptical trajectory, moving away from the BH, where the gravitational field is relatively weaker. This motion corresponds to quieter segments in the GWs, observed in both $h_+$ and $h_\times$ polarizations, and aligns with the leaves of the orbit. As the particle spirals inward, it enters the whirl phase,
tracing a tighter, nearly circular path in the strong-field region near the BH. This results in intense bursts in the GW signal due to the enhanced gravitational interaction. The number of quiet intervals corresponds to the number of leaves, while the louder bursts reflect the number of whirls; both features directly encode the orbital structure defined by the integer triplet.

In Figs.~\ref{Fig(-0.05)}-\ref{Fig(411)}, the GW strains $h_+$ and $h_\times$ are shown normalized by $10^{-21}$. Masses are converted from geometrized to SI units using $M\rightarrow\frac{GM}{c^2}$, introducing the overall factor $\frac{G^2}{c^4}$, and the luminosity distance $D_L$ is converted from $Mpc$ to meters $(1Mpc=3.085677581\times10^{22}m)$ with $G=6.67430\times10^{-11}m^3kg^{-1}s^{-2}$ and $c=2.99792458\times10^8 m/s$.
\section{Fundamental Frequencies \label{section freq.}}
The angular velocity / Keplerian frequency of the test particle, measured by a distant observer, is given by \cite{galaxies11060113}
\begin{equation}
\label{Omega}
\Omega_K=\Omega_\phi=\frac{d\phi}{dt}=\frac{\dot{\phi}}{\dot{t}} \ . 
\end{equation}
For line element \eqref{1}, the expression for the Keplerian frequency becomes
\begin{equation}
   \Omega_{\phi}=
   \frac{\sqrt{M} e^{-\frac{r}{2 \lambda }} \sqrt{\lambda  \left(\delta +e^{r/\lambda }\right)+\delta  r}}{\sqrt{\delta +1} \sqrt{\lambda } r^{3/2}}.
\end{equation}
afterward\begin{align}
    \Omega_\theta^2(r_0) &= \Gamma^\theta_{\phi\phi,\theta} \left(u^\phi\right)^2, \nonumber \\
    \Omega_r^2(r_0) &= \left( \Gamma^r_{jl,r} - 4 \Gamma^r_{jk} \Gamma^k_{rl} \right) u^j u^l, \quad (j, k, l = t, \phi). \label{alternate}
\end{align}
For spherically symmetric, non-rotating spacetimes, the expressions for $\Omega_r^2(r_0)$ can be expanded as
\begin{align}
    \Omega_r^2(r_0) &= \left(\Gamma^r_{tt,r} - 4 \Gamma^r_{tt} \Gamma^t_{rt}\right) \left(u^t\right)^2 + \left(\Gamma^r_{\phi\phi,r} - 4 \Gamma^r_{\phi\phi} \Gamma^\phi_{r\phi}\right) \left(u^\phi\right)^2.
\end{align}
To express the fundamental frequencies $\nu_r$ and $\nu_\theta$ in terms of the coordinate time $t$, we use 
\begin{align}
    \nu_i &= \frac{\Omega_i}{2\pi} \frac{d\tau}{dt}, \label{1.5.26} (i=r,\theta).
\end{align}
For line element \eqref{1}, we can write the fundamental frequencies of the radial and vertical oscillation of the test particle moving along the circular orbits around an ENLMY BH as

\begin{strip}
\begin{equation}
\left(\frac{\nu_{r}}{\nu_{\phi}} \right)^2=
\frac{
(\delta +1) r e^{r/\lambda} \left(
 \lambda^{2} e^{\frac{2 r}{\lambda}} (r-6 M + \delta r-2 \delta^{2} M (3 \lambda^{2} + r^{2} + 5 \lambda r)
- \delta e^{r/\lambda} \bigl( (\delta +1) r (-\lambda^{2} + r^{2} - \lambda r) - 2 M (r - 6 \lambda)(\lambda + r) \bigr) )
\right)
}{
\lambda \left(\lambda (\delta + e^{r/\lambda}) + \delta r \right)
\left(
-4 \delta^{2} M^{2} + e^{\frac{2 r}{\lambda}} \left((\delta +1)^{2} r^{2} - 4 M^{2}\right) - 8 \delta M^{2} e^{r/\lambda}
\right)
}
 ,
\end{equation}
\end{strip}
and
\begin{equation}
    \nu_{\theta}=\nu_{\phi}=\frac{1}{2\pi}\frac{\sqrt{M} e^{-\frac{r}{2 \lambda }} \sqrt{\lambda  \left(\delta +e^{r/\lambda }\right)+\delta  r}}{\sqrt{\delta +1} \sqrt{\lambda } r^{3/2}},
\end{equation}
where $\nu_{\phi}=\frac{\Omega_{\phi}}{2\pi}$. We can observe that $\nu_{\theta}=\nu_{\phi}$ as the spacetime is non-rotating. 
\begin{figure}
   \centering
\includegraphics[width=0.87\linewidth]{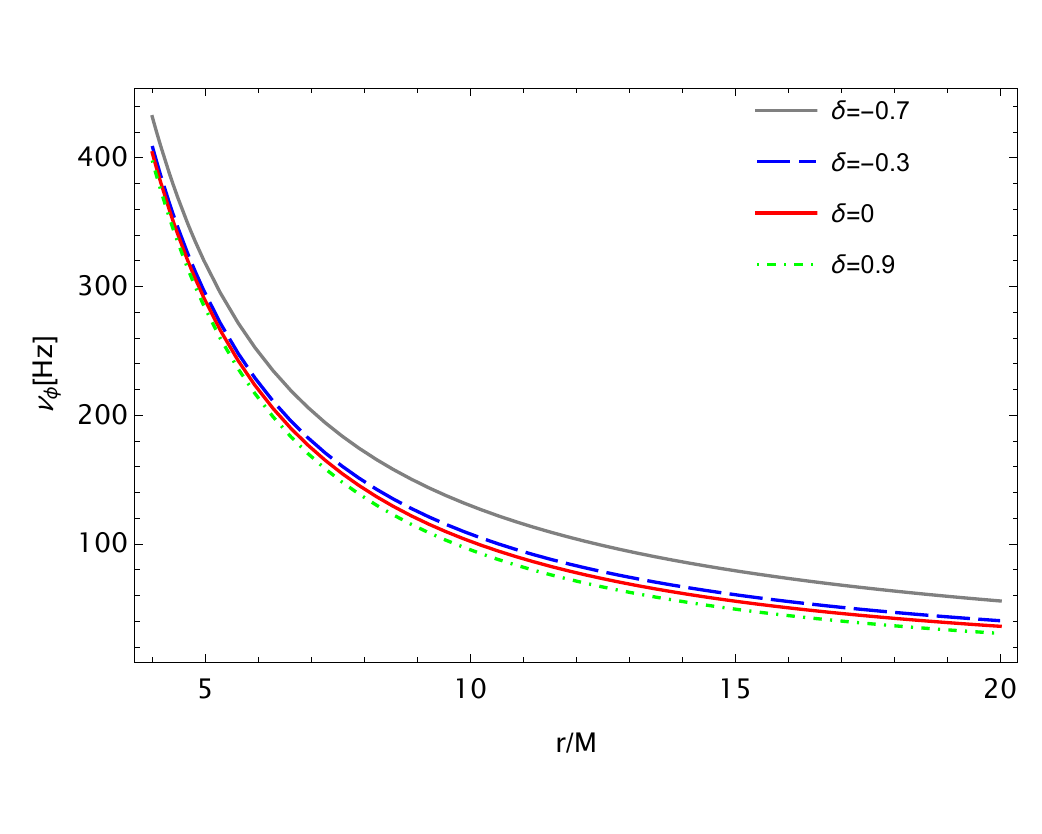} 
	\caption{$\nu_{\phi}$ is plotted for different values of $\delta$ and $\lambda=10$ by considering an object with mass $M = 10M_\odot$. The Schwarzschild case corresponding to \(\delta = 0\) is given in red.}\label{keplerFrequency}
\end{figure}
\begin{figure}
   \centering
\includegraphics[width=0.87\linewidth]{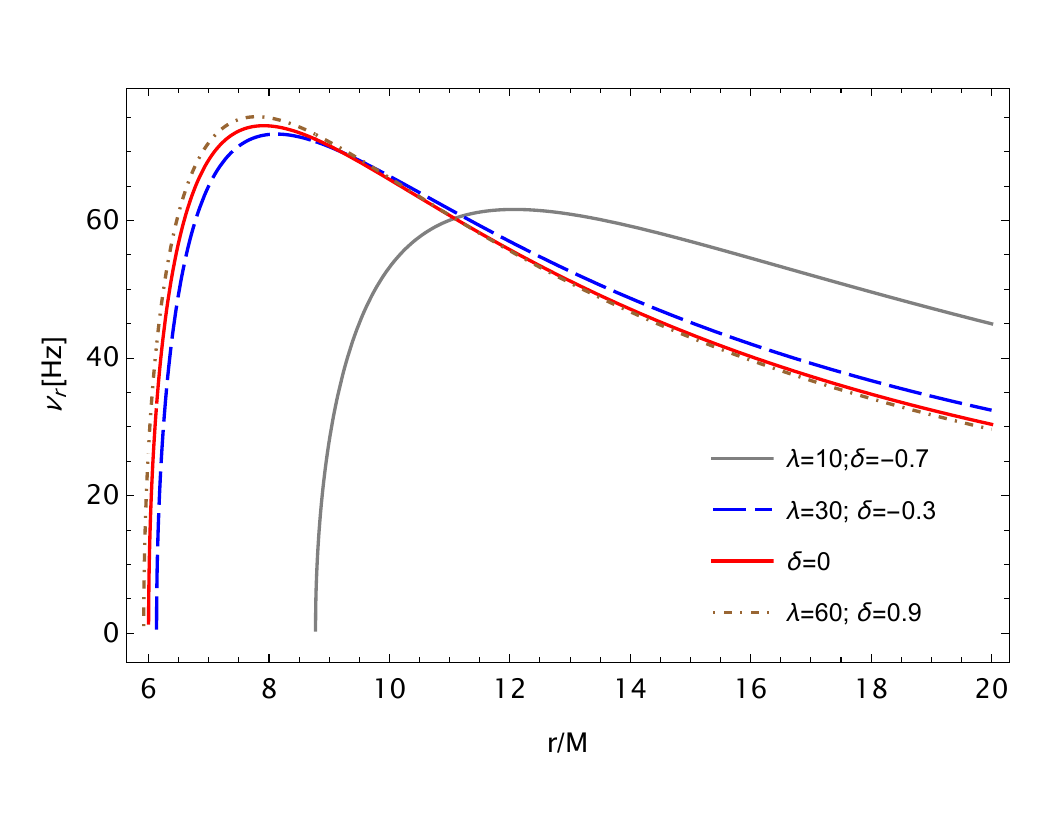} 
	\caption{We have considered an object with mass $M = 10M_\odot$. $\nu_{r}$ are plotted for different values of $\delta$ and $\lambda$. We can observe that it vanishes at $r=r_{\text{ISCO}}$. The Schwarzschild case corresponding to \(\delta = 0\) is given in red.}
    \label{NU_r}
\end{figure}
\par
In Fig.~\ref {keplerFrequency} and \ref{NU_r}, we have plotted the fundamental frequencies against $r$ after converting the geometrical units into Hertz by using the factor $\frac{c^3}{GM}$, where the speed of light \( c = 3 \times 10^{10} \, \mathrm{cm/s} \) and the gravitational constant \( G = 6.67 \times 10^{-8} \, \mathrm{cm^3 \, g^{-1} \, s^{-2}} \) \cite{RAYIMBAEV2022100930,Jusufi:2020odz}. We can note that $\nu_{r}|_{r=r_{\text{ISCO}}}=0$, indicating that for \(r < r_{\text{ISCO}}\), radial perturbations no longer produce oscillations but instead cause the orbit to become unstable.
\subsection{Quasi-Periodic Oscillation Frequencies}
In this subsection, we analyze the behavior of twin-peak QPO frequencies around an ENLMY BH, compare them with the Schwarzschild case, and discuss the relationship between the possible upper $(\nu_U)$ and lower $(\nu_L)$ QPO frequencies within three theoretical frameworks: the RP model ( $\nu_U=\nu_\phi$ and $\nu_L=\nu_\phi-\nu_r$), the WD model ($\nu_U = 2\nu_K - \nu_r$ and $\nu_L = 2(\nu_K - \nu_r)$), and the TD model ( $\nu_U = \nu_K + \nu_r$ and $\nu_L = \nu_K$) \cite{Kotrlova:kerr-naked,model1,model2,Stuchl_k_2016}. By comparing predictions for  ENLMY and Schwarzschild BHs, we identify characteristic signatures that may distinguish between standard and modified gravity scenarios.
\begin{figure*}
   \centering
\includegraphics[width=0.4\linewidth]{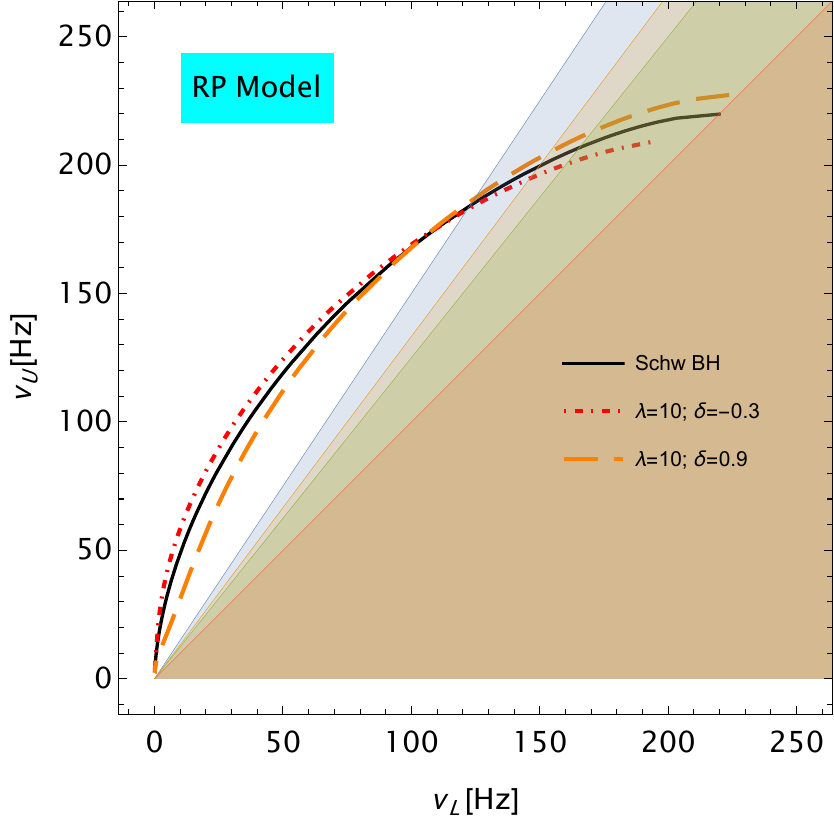}
\includegraphics[width=0.4\linewidth]{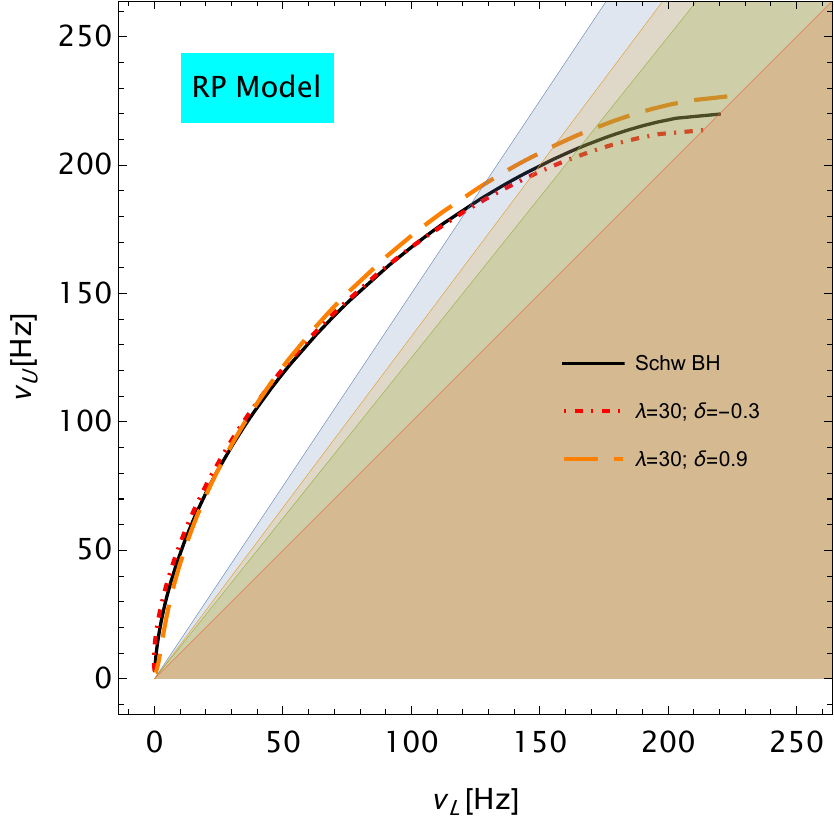}
\includegraphics[width=0.4\linewidth]{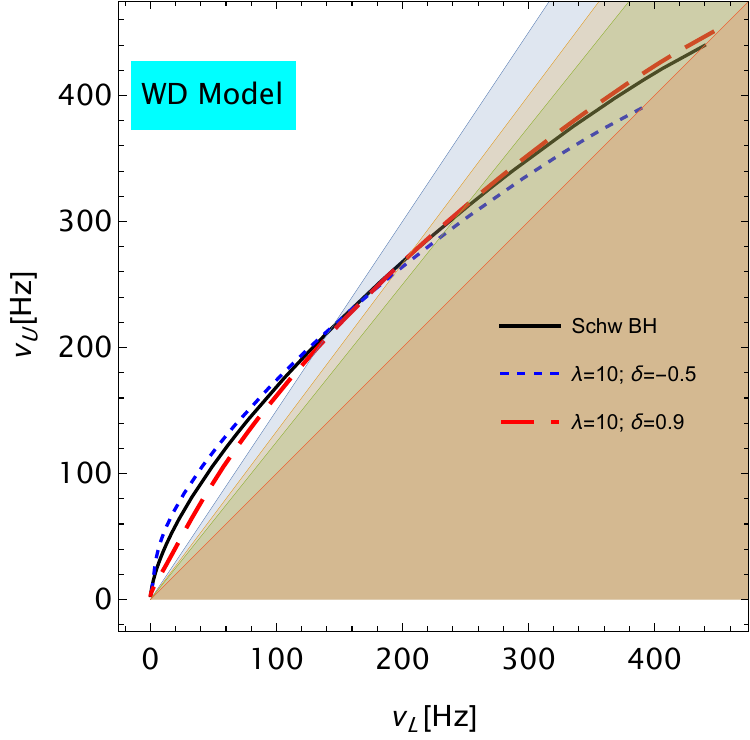}
\includegraphics[width=0.4\linewidth]{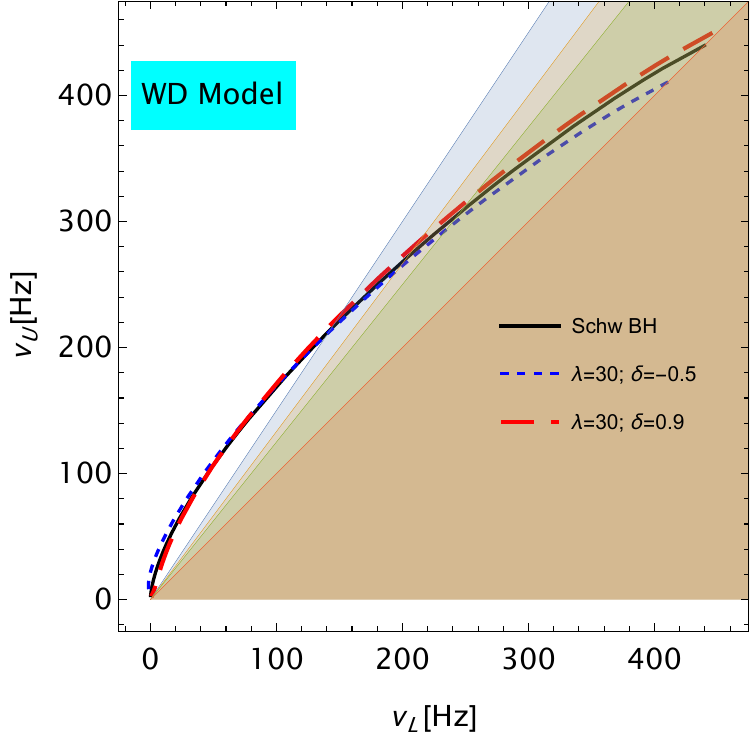}
\includegraphics[width=0.4\linewidth]{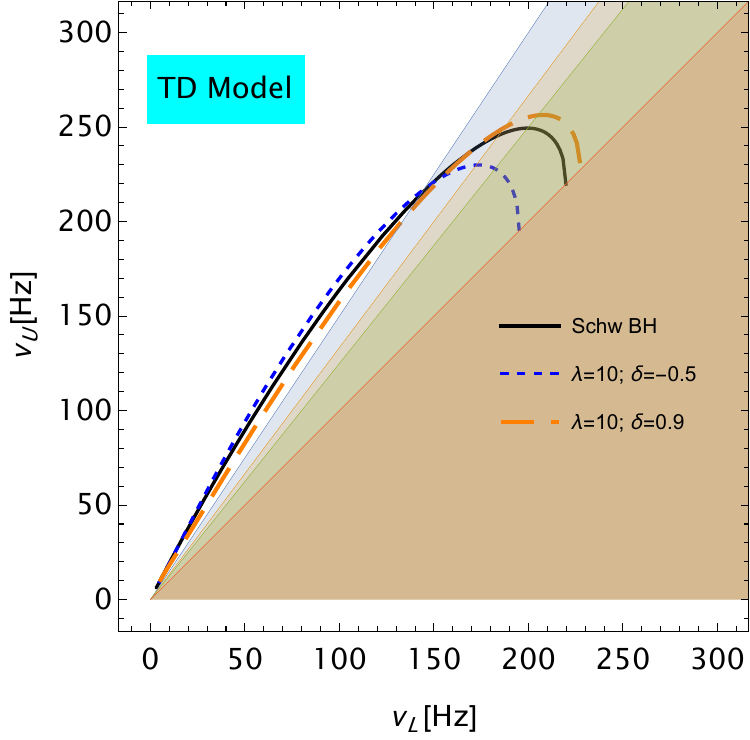}
\includegraphics[width=0.4\linewidth]{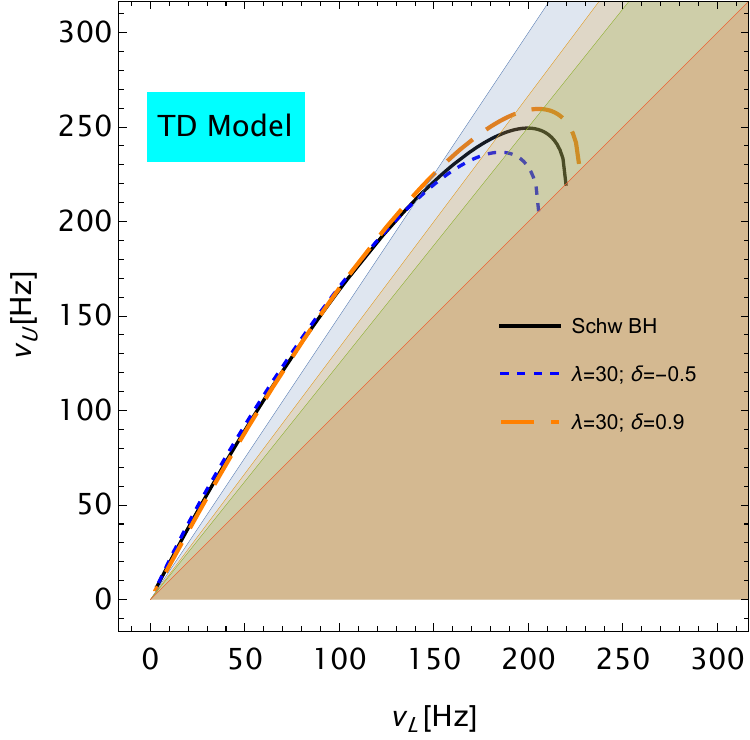}
\caption{Relations between the frequencies of upper and lower picks of twin-pick QPOs in the RP, WD, and TD models.} \label{QPO}
\end{figure*}

Fig.~\ref{QPO}, presents our results, showing the relationship between $(\nu_U)$ and $(\nu_L)$ QPO frequencies for each model under different parameters values of $\lambda$ and $\delta$. 
For numerical calculations, we fix the BH mass at $M = 10\,M_\odot$. In each model, the region between the  ENLMY and Schwarzschild curves represents possible observational outcomes for twin-peak QPOs. For all three models considered, we observe that at a fixed value of \(\lambda = 10\), the curves tend to converge within a certain intermediate region of the radial coordinate. In this region, the influence of the deformation parameter becomes negligible, and the models closely mimic the behavior of the Schwarzschild geometry. This ``mimic region'' is a common feature across all models and reflects a zone where modifications to GR are effectively suppressed. Furthermore, when \(\lambda\) is increased to \(\lambda = 30\), we find that this mimic region becomes more extended. This indicates that a larger characteristic length scale \(\lambda\) allows the exponential corrections to decay more gradually, resulting in the models resembling Schwarzschild behavior over a wider radial domain.

Yukawa corrections are most relevant in strong fields, where deviations from GR are observed in QPO signals. Micro-quasars and galactic centers are therefore an alternative to probe ENLMY gravity through high-frequency oscillations.

\section{MCMC Analyses}\label{section6}

This section is devoted to obtaining constraints on the BH mass and Yukawa parameters $\delta$ and $\lambda$. Here, we select three different BH candidates: stellar-mass, intermediate-mass, and supermassive BHs in different objects, such as stellar mass BHs, which are located at the center of the microquasars GRO J1655-40, XTE 1550-564, and GRS 1915+105. We also used QPO data from M82 X-1 \cite{pasham2014blackhole,2015MNRAS.451.2575S}. QPO data from an intermediate-mass BH is at the ultraluminous X-ray source in galaxy M82. Its mass is estimated to be around 400 to 900 solar masses in the literature \cite{2008MNRAS.387.1707C,2006ApJ...637L..21D,2006MNRAS.365.1123M,2010ApJ...712L.169F,2006MNRAS.370L...6P,2010ApJ...710...16Z}. Also, a supermassive BH Sgr A* is also in our focus with milliHertz QPOs. The QPO data of the objects are given in Table \ref{table1}.

\begin{table*}[ht!]\centering
\begin{center}
\caption[]{\label{table1} QPOs frequencies observed in microquasars and Galactic centre}
\renewcommand{\arraystretch}{1.2}
\begin{tabular}{| l || c  c | c  c | l |}
\hline
Source 
&  $\nu_{\rm{U}}\,$[Hz]&$\Delta\nu_{\mathrm{U}}\,$[Hz]& $\nu_{\rm {L}}\,$[Hz]&$\Delta\nu_{\rm{L}}\,$[Hz]& Mass [\,M$_{\odot}$\,] \\
\hline
\hline
XTE~J1550-564 \cite{2011MNRAS.416..941S} & 276&$\pm\,3$& 184&$\pm\,5$&  ${12.4^{+2.0}_{-1.8}}$\\
GRO~J1655--40 \cite{Strohmayer2001}    & 451&$\pm\,5$& 298&$\pm\,4$ & {5.4$\pm$0.3}   \\
GRS~1915+105 \cite{2015ApJ...814...87M}   & 168&$\pm\,3$& 113&$\pm\,5$&  ${12.4^{+2.0}_{-1.8}}$   \\
M82 X-1 \cite{pasham2014blackhole}  & 5.07 & $\pm\,0.06$ & 3.32& $\pm\,0.06$ &$415\pm\,63$\\
Sgr~A* \cite{Ghez:2008}  & 1.445&$\pm 0.16$ mHz  &0.886&$\pm 0.04$ mHz  & (4.1 $\pm$ 0.6)\,$\times 10^6$\\ 

\hline
        \end{tabular}
\end{center}
\end{table*}

To obtain the estimation for the four parameters as the QPOs frequencies observed in the microquasars, we perform the $\chi^2$ analysis with \cite{Bambi2013arXiv1312.2228B}:
\begin{eqnarray}\nonumber
\chi^{2}\left(M,\delta,\lambda,r\right)&=&\frac{(\nu_{1\phi}-\nu_{1\rm U})^{2}%
}{\sigma_{1\rm U}^2}+\frac{(\nu_{1\rm per}-\nu_{1\rm L})^{2}%
}{\sigma_{1 \rm L}^2}\\ \nonumber
&+&\frac{(\nu_{1\rm nod}-\nu_{1\rm C})^{2}%
}{\sigma_{1\rm C}^2}+\frac{(\nu_{2\phi}-\nu_{2\rm U})^{2}}%
{\sigma_{2\rm U}^2}\\
&+&\frac{(\nu_{2\rm nod}-\nu_{2\rm C})^{2}}%
{\sigma_{2\rm C}^2}~.
\end{eqnarray}

The best-estimated values of the parameters $M$, $\delta$, $\lambda$, and $r$ (radius of QPO orbits) in which $\chi_{\rm min}^{2}$ take minimum. The range of the parameters at the confidence level (C.L.) can be determined in the interval $\chi_{\rm min}^{2}\pm \Delta\chi^{2}$. To find these best-estimated values, we minimize the $\chi^2$ and get the following mean values for the $M$, $\delta$, $\lambda$, and $r$ in Table \ref{prior}.

\begin{table*}[ht!]\centering
\begin{center}
\renewcommand\arraystretch{1.5} 
\caption{\label{prior}%
The Gaussian prior of BH spacetime in EU from QPOs for the microquasars and galactic centers.}
\begin{tabular}{lcccccccccc}
\hline\hline
\multirow{2}{*}{$P$} & \multicolumn{2}{c}{XTE J1550-564} & \multicolumn{2}{c}{GRO J1655-40} & \multicolumn{2}{c}{GRS 1915+105} &\multicolumn{2}{c}{Sgr $A^*$} & \multicolumn{2}{c}{M82 X-1} \\
& $\mu$ & \multicolumn{1}{c}{$\sigma$} & $\mu$          & $\sigma$ & $\mu$ & \multicolumn{1}{c}{$\sigma$} & $\mu$ & \multicolumn{1}{c}{$\sigma$} & $\mu$ & \multicolumn{1}{c}{$\sigma$} \\
\hline
     $M/M_{\odot}$ & $12.09$ &1.03 & $5.39$ & $0.17$ & $12.51$ & $1.10$ & $3.82\times10^6$ & $0.25\times10^6$ & $416.23$ & $22.79$\\
    
     $\delta$ & 0.66 & 0.24   & $0.61$  & $0.24$ & 0.61 & 0.24 & $0.83$ & $0.13$ & $0.61$ & $0.23$ \\ 
    
     $\lambda$ & 0.46 & 0.29 & $0.61$ & $0.41$ & $0.59$ & $0.39$ & 0.21 & 0.09 & 0.61 & 0.39\\
    
     $r$ & 4.03 & 0.38 & $4.69$ & $0.37$ & 5.16 & 0.39 & 3.44 & 0.24 & 5.04 & 0.39\\
     \hline\hline
\end{tabular}
\end{center}
\end{table*}

We used the Python library \texttt{emcee} \cite{emcee,JR.MCMC2024} to perform the MCMC analysis and constrain the parameters of the particle around an ENLMY BH. Our analysis used the relativistic precision (the RP see Section \ref{section freq.}) method.

The posterior distribution can be defined as follows \cite{Liu-etal2023},
\begin{eqnarray}
\mathcal{P}(\theta |\mathcal{D},\mathcal{M})=\frac{P(\mathcal{D}|\theta,\mathcal{M})\ \pi (\theta|\mathcal{M})}{P(\mathcal{D}|\mathcal{M})},
\end{eqnarray}
where $\pi(\theta)$ is the prior and $P(D|\theta,M)$ is the likelihood. We choose our priors to be (normal) Gaussian within suitable boundaries (see Table \ref{prior}), \textit{i.e.,} 
$\pi(\theta_i) \sim \exp\left[{\frac{1}{2}\left(\frac{\theta_i - \theta_{0,i}}{\sigma_i}\right)^2}\right]$, \ $\theta_{\text{low},i}<\theta_i<\theta_{\text{high},i}$. In this work, the parameters are $\theta_i=\{M,\delta,\lambda,r/M\}$ and $\sigma_i$ are their corresponding variances. Following the orbital, periastron precession frequencies from Sect.\ref{section freq.}, different parts of the data are used in our MCMC analysis. Eventually, the likelihood function $\Lambda$ can be expressed as
\begin{eqnarray}
\log \Lambda = \log \Lambda_{\rm U} + \log \Lambda_{\rm L},\label{likelyhood}
\end{eqnarray}
where $\log \Lambda_{\rm U}$ denotes the likelihood of the astrometric upper/orbital frequencies,
\begin{eqnarray}
 \log \Lambda_{\rm U} = - \frac{1}{2} \sum_{i} \frac{\left(\nu_{\phi\rm, obs}^i -\nu_{\phi\rm, th}^i\right)^2}{\left(\sigma^i_{\phi,{\rm obs}}\right)^2} \ ,
\end{eqnarray}
and $\log \Lambda_{\rm L}$ represents the probability (likelihood) of the data of the lower or periastron precession frequency ($\nu_{\rm per}$).

\begin{eqnarray}
\log \Lambda_{\rm L} =-\frac{1}{2} \sum_{i} \frac{\left(\nu_{\rm per, obs}^i -\nu_{\rm per, th}^i\right)^2}{\left(\sigma^i_{\rm per,{\rm obs}}\right)^2}.\ 
\end{eqnarray}
Here $\nu^i_{\phi,\rm obs}$, $\nu^i_{\rm per,\rm obs}$ are observational results of the orbital/Keplerian frequencies ($\nu_{\rm K}$), periastron precession frequencies $\nu_{\text{per}}=\nu_{\rm K}-\nu_{\rm r}$ for the sources mentioned above. On the other side, $\nu^i_{\phi,\rm th}$, $\nu^i_{\rm per,\rm th}$ are the respective theoretical estimations.

Next, we perform the MCMC simulation to constrain the parameters ($M$, $\delta$, $\lambda$, $r/M$) for an ENLMY BH. We use Gaussian priors based on parameter values from the existing literature on QPO data processing. We sampled approximately $10^5$ points for each parameter from a Gaussian prior distribution, allowing us to explore the physically possible parameter space within set boundaries and identify the best-fitting parameter values.
\begin{figure}[H] \centering
\includegraphics[width=0.8\linewidth]{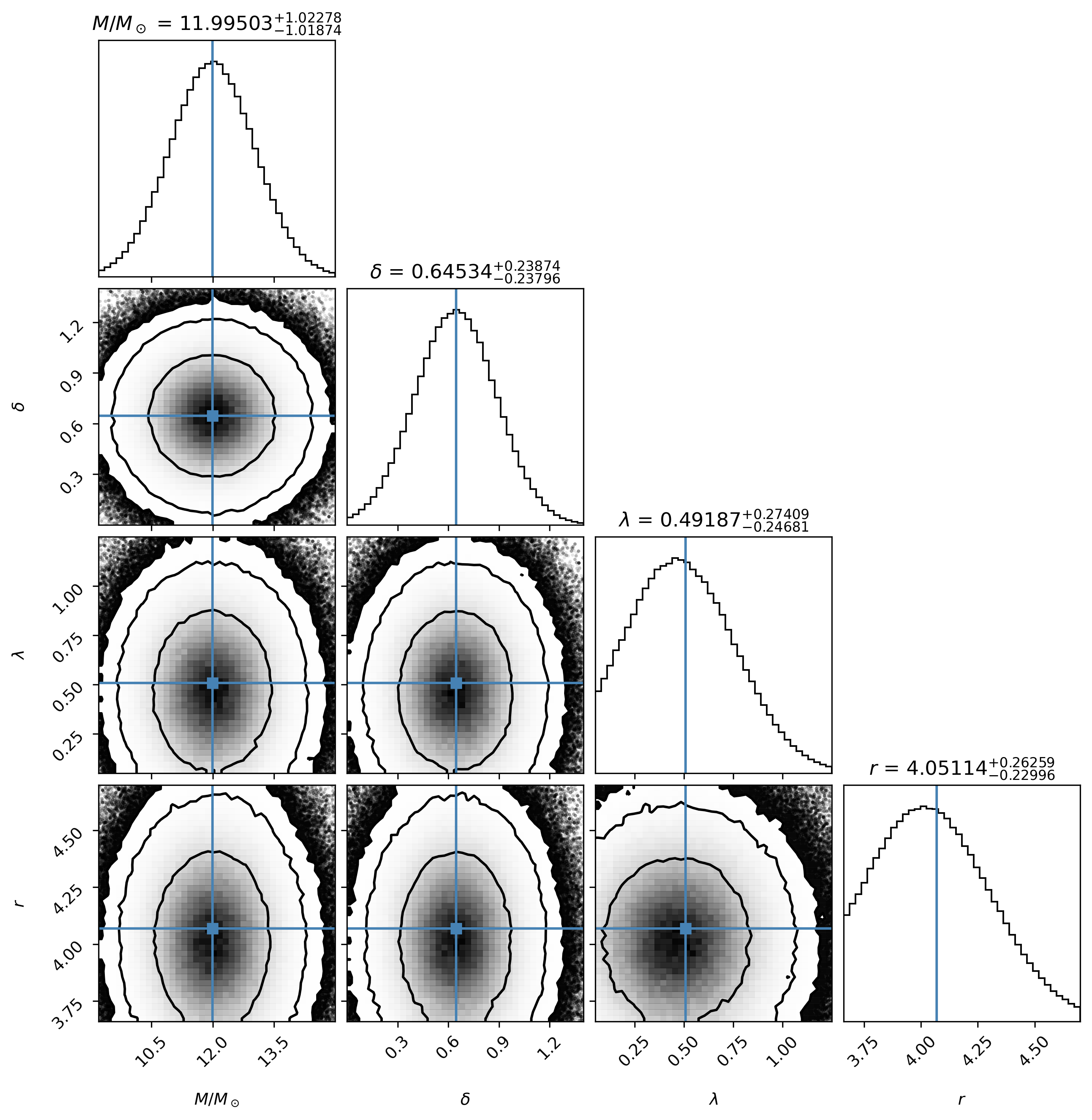}
\caption{Constraints on the Yukawa parameters, the BH mass, and the radius of the QPO orbit from a four-dimensional MCMC analysis using the QPO data for the stellar-mass BH XTE J1550-564 in the RP model. \label{XTE}} \end{figure}


\begin{figure*} \centering
\includegraphics[width=0.45\linewidth]{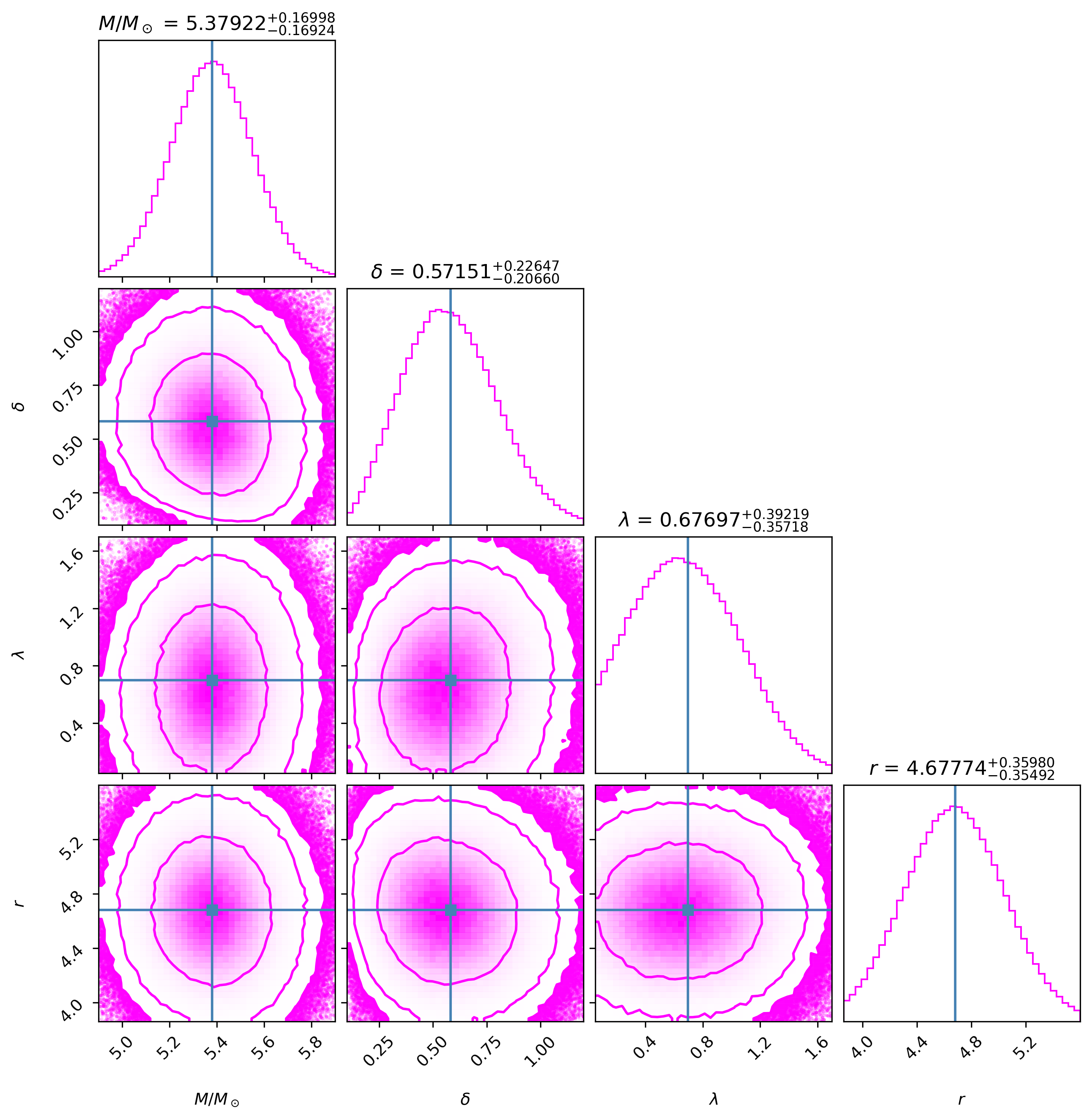}
\includegraphics[width=0.45\linewidth]{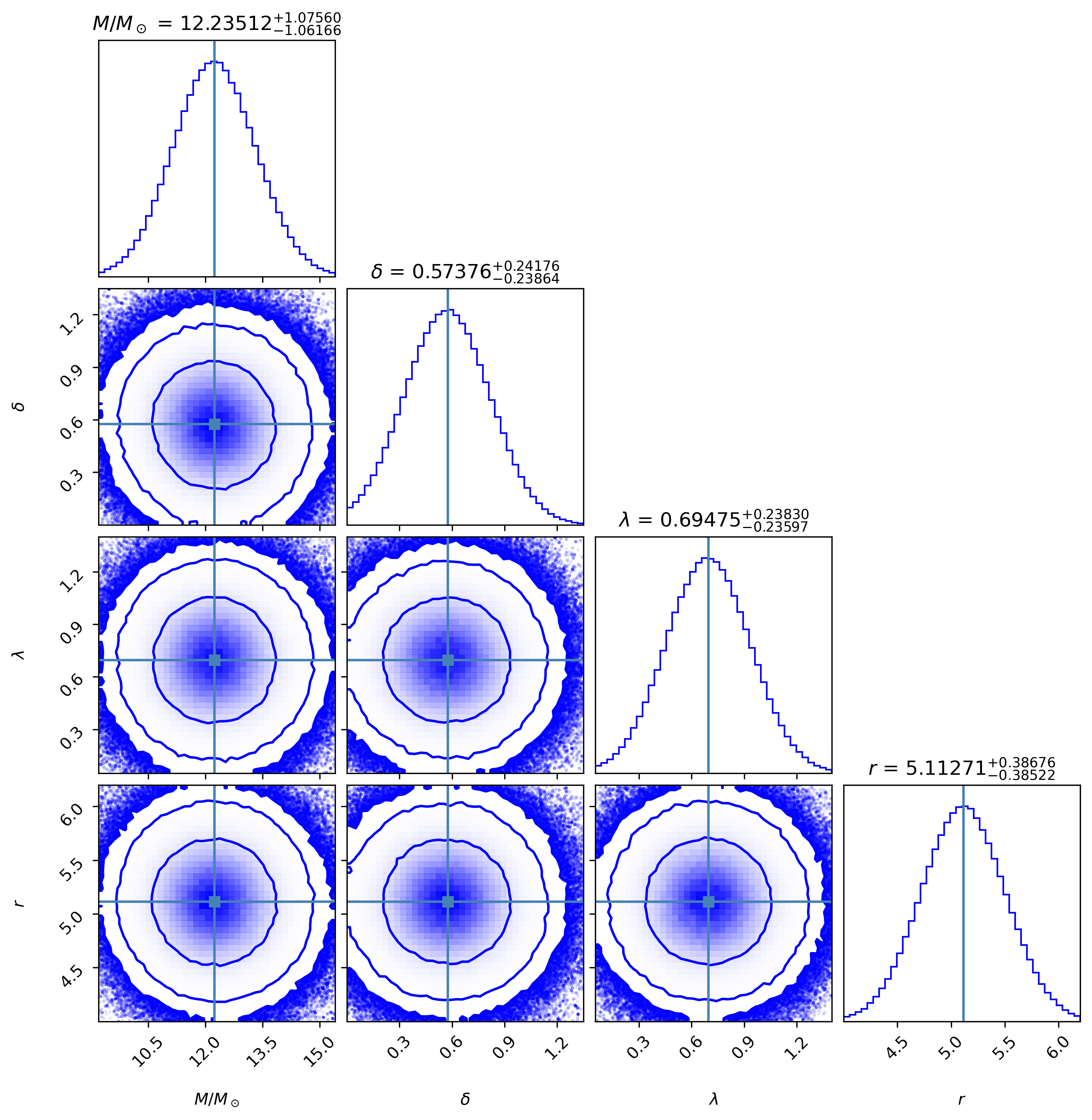}
\includegraphics[width=0.45\linewidth]{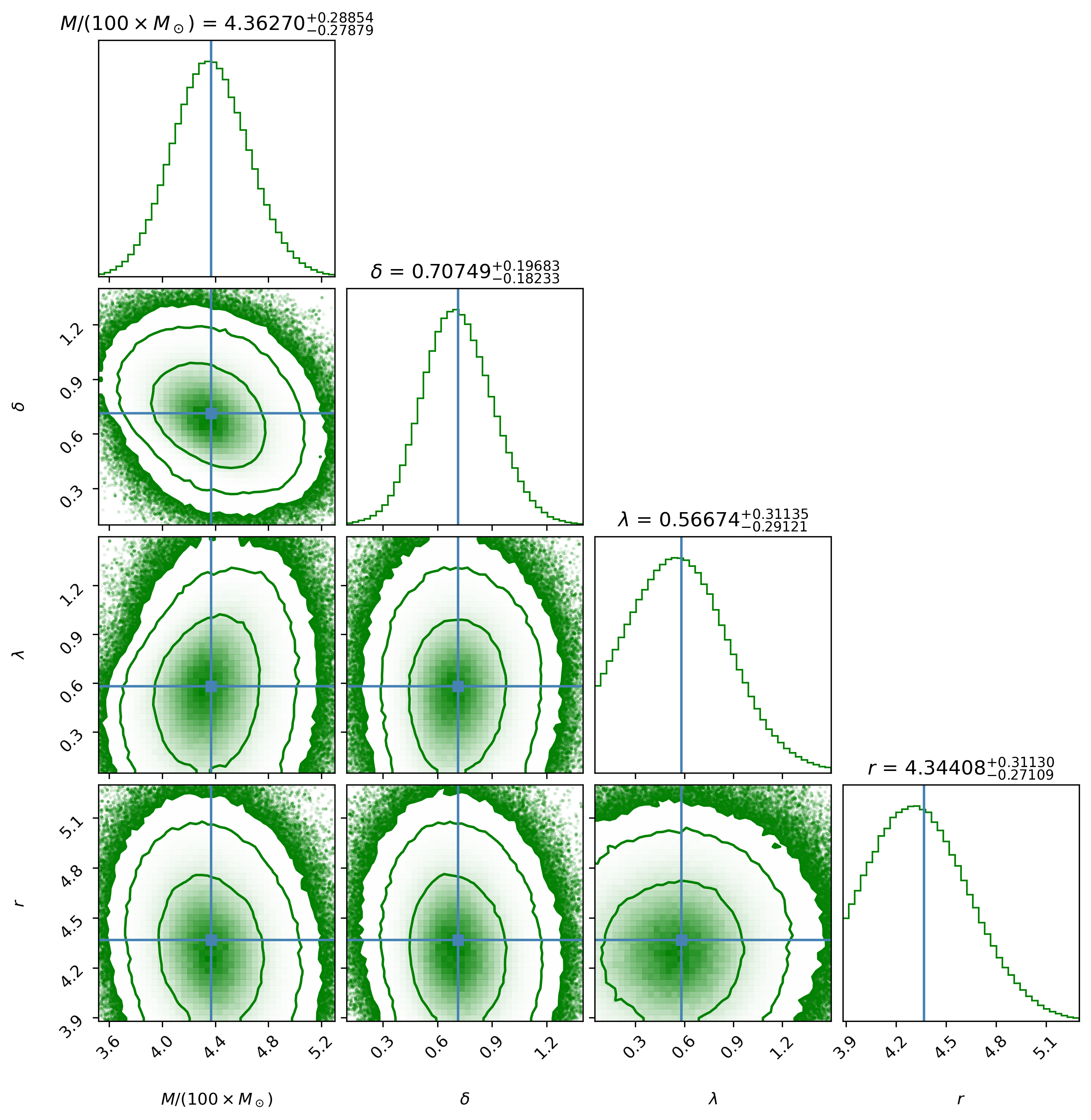}
\includegraphics[width=0.45\linewidth]{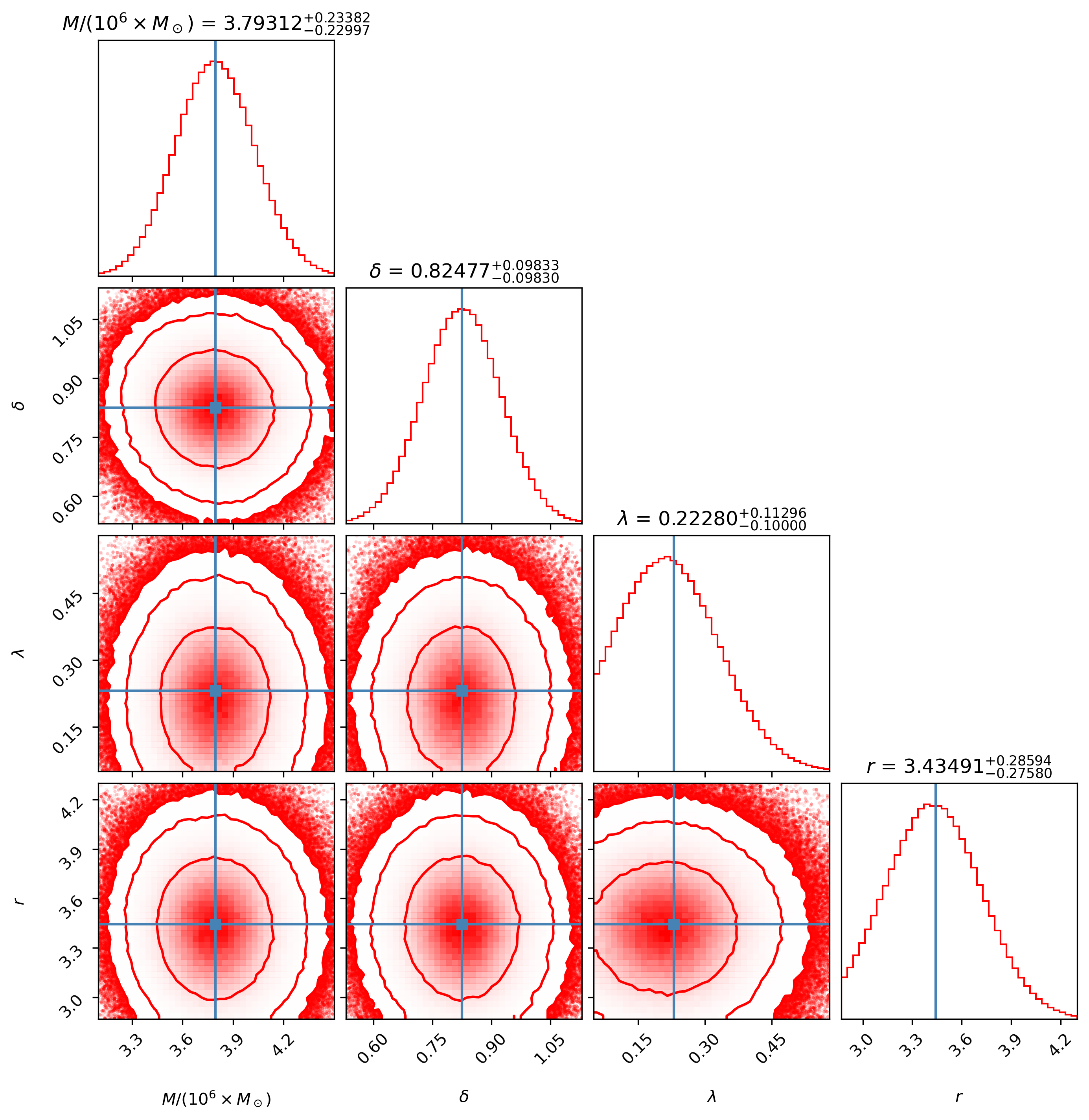}
\caption{The same figure with Fig.\ref{XTE}, but for GRO J1655-40 (in the top-left panel with magenta) and GRS 1915+105 (in the top-right panel with blue), M82 X-1 (in the bottom-left panel with green) and Sgr A$^*$ (in the bottom-right panel with red). \label{XTESgrA}} \end{figure*}

Now, we collect the best-fit values for the parameters $(M,\delta,\lambda,r)$ obtained from the MCMC results in Figs. \ref{XTE} and \ref{XTESgrA} and present them in a table form.

\begin{table*} \centering \renewcommand\arraystretch{1.5} \caption{The best-fit values of the BH in ENLMY spacetime.} \label{tab:best_fit_value} 
\begin{tabular}{lccccc} \hline\hline \multicolumn{1}{c}{$P$} & \multicolumn{1}{c}{XTE J1550-564} & \multicolumn{1}{c}{GRO J1655-40} & \multicolumn{1}{c}{GRS 1915+105} & \multicolumn{1}{c}{M82 X-1} & \multicolumn{1}{c}{Sgr $A^*$} \\ \hline 
$M/M_{\odot}$ & $11.995^{+1.028}_{-1.018}$ & $5.379^{+0.169}_{-0.169}$ & $12.235^{+1.075}_{-1.061}$ & $4.362^{+0.288}_{-0.278} \times 10^2$ & $3.793^{+0.233}_{-0.229} \times 10^6$ \\ 
$\delta$ & $0.645^{+0.238}_{-0.237}$ & $0.571^{+0.226}_{-0.206}$ & $0.573^{+0.241}_{-0.238}$ & $0.707^{+0.196}_{-0.182}$ & $0.824^{+0.098}_{-0.098}$ \\ 
$\lambda$ & $0.491^{+0.274}_{-0.229}$ & $0.676^{+0.392}_{-0.357}$ & $0.694^{+0.238}_{-0.235}$ & $0.566^{+0.311}_{-0.291}$ & $0.223^{+0.112}_{-0.100}$\\ 
$r/M$ & $4.052^{+0.268}_{-0.230}$ & $4.677^{+0.392}_{-0.357}$ & $5.112^{+0.386}_{-0.385}$ & $4.344^{+0.311}_{-0.271}$ & $3.435^{+0.285}_{-0.275}$ \\ \hline\hline \end{tabular} \end{table*}
 
Our estimated masses exhibit minor deviations relative to the limits reported in our earlier study as well as those presented \cite{2025EPJC...85..126J,2024EPJC...84..964J,2024ChJPh..92..143R,2024EPJC...84.1114R,2023Univ....9..391M,2023Galax..11..113R,2024ChPhC..48e5104R,2024JHEAp..44...99C,2024PDU....4601708M,2024PDU....4601561M}. The oscillation frequencies depend on not only the BH mass but also gravity parameters, in particular, the Yukawa $\delta$ and $\lambda$ parameters. Therefore, the results of the constraint with the QPO frequencies should also differ for each case.

\section{Conclusion}
In this paper, we investigated the motion of test particles around the ENLMY BH, a modified spacetime arising from $f(R)$  gravity coupled with non-linear electrodynamics and Yukawa-type corrections. Beginning with the analysis of the spacetime metric and its associated symmetries, we derived the equations of motion using the Hamiltonian formalism. By exploiting the conserved quantities associated with Killing vectors, we investigated the dynamics of particles confined to the equatorial plane, enabling us to simplify and understand the behavior of geodesic motion under the influence of the Yukawa potential.

Our analysis of the effective potential revealed critical insights into the structure of circular orbits. We derived expressions for the specific energy and angular momentum of particles in stable circular motion and identified the locations of the ISCO and IBCO. These orbits are sensitive to the Yukawa parameters $\lambda$ and $\delta$, with increasing values that shift the ISCO and IBCO to larger radii, requiring higher energy and angular momentum for circular motion. Through numerical analysis, we showed how these parameters affect the structure of bound orbits and presented detailed profiles of the effective potential and its dependence on the spacetime deformation.

Furthermore, we explored periodic orbits using Levin’s classification scheme \cite{levin2008periodic,levin2009energy}, characterizing them by the rational number $q$, which depends on the energy, angular momentum, and spacetime parameters $\lambda$ and $\delta$. Our results reveal that the orbital dynamics around the ENLMY BH are strongly influenced by the Yukawa parameters $\lambda$ and $\delta$. For small values of $\lambda$, the energy required for periodic orbits in the ENLMY spacetime is generally higher than that in the Schwarzschild case, regardless of the sign of $\delta$. However, as $\lambda$ increases, the energy of periodic orbits becomes lower than that of the Schwarzschild counterpart for both positive and negative $\delta$. This behavior indicates that Yukawa corrections can either weaken or strengthen the gravitational binding depending on the parameter regime, thereby modifying the conditions for closed orbits and enriching the structure of the orbital phase space.

Additionally, we analyzed the GW radiation emitted by periodic orbits characterized by different $(z,w,v)$ configurations in the ENLMY spacetime. The resulting waveforms for the plus $(h_+)$ and cross $(h_\times)$ polarizations display characteristic zoom-whirl structures, as illustrated in Figs.~\ref{Fig(-0.05)}-\ref{Fig(411)}. Specifically, the waveforms exhibit quiet phases during the extended zoom segments of the orbit and sharp, high-amplitude glitches during the rapid whirl phases. The number of these glitches corresponds directly to the number of whirls in each orbit, while the quiet intervals align with the orbital leaves. These waveform patterns reflect the impact of the Yukawa parameters $\lambda$ and $\delta$ on the GW phase evolution, indicating how modifications to the gravitational potential affect orbital dynamics. Such distinctive features in the GW signal offer promising observational signatures that may serve to test and constrain the ENLMY BH model with upcoming GW detectors.
\par
We determined the fundamental frequencies of a test particle using the perturbed geodesic equation in the vicinity of a circular orbit. We observed that as $r$ decreases, the radial fundamental frequency increases, attains a certain peak, and then starts decreasing. Eventually, it vanishes at the radius of ISCO. We have also analyzed the twin-peak QPO frequencies for three models and observed that increasing the characteristic length scale increases the mimic region. This is because a larger length scale means the deviations from GR are weaker, and hence the behavior of the BH becomes more similar to the Schwarzschild case over a wider range of radii. Hence, we can say that astrophysical observations may not easily distinguish between modified gravity and GR. Consequently, high-precision observations could be employed to place meaningful constraints on the allowed values of $\lambda$.
\par
Finally, in the last section, we have provided MCMC analyses to obtain constraints on the ENLMY BH mass parameters. We have chosen data from QPOs observed in X-ray binaries: GRO J1655-40, GRS 1915-105, and XTE J1550-564. Details of MCMC analyses are shown in Figs.\ref{XTE} \& \ref{XTESgrA}, and best values are provided in Tables \ref{tab:best_fit_value}. 

In summary, the ENLMY black hole provides a natural testing ground for probing intermediate-scale deviations from GR. By combining QPO data with gravitational wave signatures, one can place observational constraints on the Yukawa parameters. This strongly motivates future observations to distinguish ENLMY gravity from other beyond-GR frameworks.

\bibliographystyle{elsarticle-num}
\bibliography{bibliography}
\end{document}